\RequirePackage[orthodox,l2tabu]{nag}

\documentclass[onecolumn]{IEEEtran}

\usepackage[utf8]{inputenc}
\usepackage[T1]{fontenc}
\usepackage[british]{babel}
\usepackage{zi4}
\usepackage{booktabs}

\usepackage{setspace}
\usepackage{authblk}

\usepackage{amsmath}
\usepackage{amsfonts}
\usepackage{amssymb}
\usepackage{amsthm}

\usepackage{varioref}
\usepackage{url}
\usepackage{cleveref}
\usepackage{cite}

\usepackage{mathtools}

\usepackage{graphicx}
\usepackage{tikz}
\usepackage{xcolor}

\theoremstyle{definition}
\newtheorem{theorem}{Theorem}[section]

\newtheorem{corollary}{Corollary}[theorem]
\newtheorem{definition}{Definition}[section]
\newtheorem{example}{Example}[section]
\newtheorem{lemma}[theorem]{Lemma}
\newtheorem{remark}{Remark}[section]

\newcommand*{\bad}[1]{{\textbf{\color{red!50!black}#1}}}

\newcommand*{\good}[1]{{\underline{\color{green!50!black}#1}}}

\newcommand\hlight[1]{\tikz[overlay, remember picture,baseline=-\the\dimexpr\fontdimen22\textfont2\relax]\node[rectangle,fill=gray!50,rounded corners,fill opacity = 0.2,draw,thick,text opacity =1] {$#1$};}

\usepackage{mdframed}

\title{From Clustering Supersequences to Entropy Minimizing Subsequences for Single and Double Deletions}

\author[1]{Arash Atashpendar\thanks{\url{arash.atashpendar@uni.lu}}}
\author[3]{Marc Beunardeau\thanks{\url{marc.beunardeau@ingenico.com}}}
\author[3]{Aisling Connolly\thanks{\url{aisling.connolly@ens.fr}}}
\author[3]{R\'emi G\'eraud\thanks{\url{remi.geraud@ens.fr}}}
\author[2]{David Mestel\thanks{\url{david.mestel@cs.ox.ac.uk}}}
\author[2]{A. W. Roscoe\thanks{\url{bill.roscoe@cs.ox.ac.uk}}}
\author[1]{Peter Y. A. Ryan\thanks{\url{peter.ryan@uni.lu}}}

\affil[1]{SnT, University of Luxembourg, Luxembourg}
\affil[2]{Department of Computer Science, University of Oxford, Oxford, UK}
\affil[3]{D\'epartement d'informatique de l'ENS, \'Ecole normale sup\'erieure, CNRS, PSL Research University, Paris, France}

\begin{document}

\maketitle

\begin{abstract}
A binary string transmitted via a memoryless i.i.d. deletion channel is received as a subsequence of the original input. From this, one obtains a posterior distribution on the channel input, corresponding to a set of candidate supersequences weighted by the number of times the received subsequence can be embedded in them. In a previous work it is conjectured on the basis of experimental data that the entropy of the posterior is minimized and maximized by the constant and the alternating strings, respectively. In this work, in addition to revisiting the entropy minimization conjecture, we also address several related combinatorial problems. We present an algorithm for counting the number of subsequence embeddings using a run-length encoding of strings. We then describe methods for clustering the space of supersequences such that the cardinality of the resulting sets depends only on the length of the received subsequence and its Hamming weight, but not its exact form. Then, we consider supersequences that contain a single embedding of a fixed subsequence, referred to as singletons, and provide a closed form expression for enumerating them using the same run-length encoding. We prove an analogous result for the minimization and maximization of the number of singletons, by the alternating and the uniform strings, respectively. Next, we prove the original minimal entropy conjecture for the special cases of single and double deletions using similar clustering techniques and the same run-length encoding, which allow us to characterize the distribution of the number of subsequence embeddings in the space of compatible supersequences to demonstrate the effect of an entropy decreasing operation.
\end{abstract}



\section{Introduction}

The original motivation for this work goes back to an analysis of quantum key distribution (QKD) protocols \cite{ryan2013enhancements}, which among other things, suggested some modifications of the quantum bit error rate (QBER) estimations. These modifications led to an information theory problem that was first investigated in \cite{atashpendar2015information}.

In QKD protocols, it is typical for the parties, after the quantum phase, to compare bits of the fresh session key at randomly sampled positions in order to obtain an estimate of the Quantum Bit Error Rate (QBER). This indicates the proportion of bits that have been flipped as the result of either noise or eavesdropping on the quantum channel. This serves to bound the amount of information leakage to any eavesdropper, and as long as this falls below an appropriate threshold, the parties continue with the information reconciliation and privacy amplification steps.

Usually, the sample set is agreed and the bits compared using un-encrypted but authenticated exchanges over a classical channel, hence the positions of the compared bits are known to a potential eavesdropper and these bits are discarded. In \cite{ryan2013enhancements}, it is suggested that the sample set be computed secretly by the parties based on prior shared secrets. They still compare the bits over an un-encrypted channel, but now an eavesdropper does not learn where the bits lie in the key stream. This prompts the possibility of retaining these bits, but now we must be careful to bound the information leakage and ensure that later privacy amplification takes account of this leakage.

The mathematical problem encountered in the aforementioned analysis is the following. A random bit string $y$ of length $n$ emitted from a memoryless source is transmitted via an i.i.d. deletion channel such that a shorter bit string $x$ of length $m$ ($m \le n$) is received as a subsequence of $y$, after having been subject to $n-m$ deletions. Consequently, the order in which the remaining bits are revealed is preserved, but the exact positions of the bits are not known. Given a subsequence $x$, the question is to find out how much information about $y$ is revealed. More specifically, the quantity we are interested in is the conditional entropy \cite{cover2012elements} over the set of candidate supersequences upon observing $x$, i.e., $H(Y|X=x)$ where $Y$ is restricted to the set of compatible supersequences as explained below.

The said information leakage is quantified as the drop in entropy \cite{shannon2001mathematical} for a fixed $x$ according to a weighted set of its compatible supersequences, referred to as the \emph{uncertainty set}. The uncertainty set, denoted by $\Upsilon_{n,x}$, contains all the supersequences that could have given rise to $x$ upon $n-m$ deletions. In \cite{atashpendar2015information}, an alternative proof shows that this set's cardinality is independent of the details of $x$ and that it is only a function of $n$ and $m$. Thus, for a fixed subsequence $x$ of length $m$, we consider the set of $y$ strings of length $n$ ($n \ge m$) that can contain $x$ as a subsequence embedding. The weight distribution used in the computation of entropy is given by the number of occurrences or embeddings of a fixed subsequence in its compatible supersequences, i.e., the number of distinct ways $x$ can be extracted from $y$ upon a fixed number of deletions, denoted by $\omega_x(y)$. Furthermore, in the same work it is conjectured that the constant subsequences consisting of all \texttt{1}'s or all \texttt{0}'s (i.e., $x= \texttt{11...1}$ and $x= \texttt{00...0}$) and the alternating \texttt{1}'s and \texttt{0}'s (i.e., $x=\texttt{1010...}$ and $x=\texttt{0101...}$), minimize and maximize the said entropy, respectively. Throughout we will simply use $\sigma$ to refer to the constant strings $x=\texttt{0}^m$ and $x=\texttt{1}^m$.

Despite the specific context in which the problem was first encountered, the underlying mathematical puzzle is a close relative of several well-known challenging problems in formal languages, DNA sequencing and coding theory. In fact, the distribution of the number of times a string $x$ appears as a subsequence of $y$, lies at the center of the long-standing problem of determining the capacity of deletion channels. More precisely, knowing this distribution would give us a maximum likelihood decoding algorithm for the deletion channel \cite{mitzenmacher2009survey}. In effect, upon receiving $x$, every set of $n-m$ symbols is equally likely to have been deleted. Thus, for a received sequence, the probability that it arose from a given codeword is proportional to the number of times it is contained as a subsequence in the originally transmitted codeword. More specifically, we have $p(y|x) = p(x|y)\frac{p(y)}{p(x)} = \omega_x(y)d^{n-m}(1-d)^m \frac{p(y)}{p(x)}$, with $d$ denoting the deletion probability. Thus, as inputs are assumed to be a priori equally likely to be sent, we restrict our analysis to $\omega_x(y)$ for simplicity.

In this work, we first study several closely-related counting problems involving (super/sub)-sequences and then we revisit the aforementioned entropy question. It is worth pointing out that while questions on the combinatorics of random subsequences requiring closed-form expressions are already quite challenging, the problem tackled in this work and first raised in
\cite{atashpendar2015information}, is further complicated by the dependence of entropy on the distribution of subsequence embeddings, i.e., the number of supersequences having specific embedding weights. To put this in contrast, in a related work \cite{swart2003note}, a closed-form expression is provided for computing the number of distinct subsequences that can be obtained from a fixed supersequence for the special case of two deletions, whereas here we need to account for the entire space of supersequences and characterize the number of times a given subsequence can be embedded in them in order to address the entropy question. Moreover, one would have to work out how these weights (number of embeddings) get shifted across their compatible supersequences when we move from one subsequence to another. To the best of our knowledge, other than the original statement of the problem \cite{atashpendar2015information} and the conjectured limiting entropic cases, proving the entropy extremization conjecture has not been addressed before.

\subsection{Results and Contributions}

We provide an algorithm for counting the number of embeddings of $x$ into $y$ as a subsequence using a run-length encoding of strings, which is used for identifying deletion patterns that simplify the counting problem to a sequential mapping of runs from $x$ strings to $y$ strings. Similar to how the cardinality of the set of supersequences that can project to a given subsequence, i.e., $|\Upsilon_{n,x}|$, depends only on their respective lengths, we prove that the number of supersequences that admit an initial embedding of a subsequence such that the last index of their initial embedding overlaps with their last bit, also depends only on $|y|=n$ and $|x|=m$. We then describe two clustering techniques that give rise to subspaces in $\Upsilon_{n,x}$ whose sizes depend only on $n, m$ and the Hamming weight of $x$, but not the exact form of $x$. We derive analytic expressions, as well as a recurrence, for the cardinality of these sets. The approach and methodology used for deriving our clustering results depend heavily on the notion of initial or canonical embeddings of subsequences in their compatible supersequences, which provide further insight into the importance of initial embeddings.

Next, we consider the problem of enumerating supersequences that admit exactly a single occurrence of a subsequence, referred to as \emph{singletons}. We provide a closed form expression for their count using the same run-length encoding and prove an analogous result for the minimization and maximization of the number of singletons, by the alternating and the uniform strings, respectively.

We then revisit the original entropy extremization question and prove the minimal entropy conjecture for the special cases of single and double deletions, i.e., for $m=n-1$ and $m=n-2$. The entropy result is obtained via a characterization of the number of strings with specific weights, along with an entropy decreasing operation. This is achieved using clustering techniques and a run-length encoding of strings: we identify groupings of supersequences with specific weights by studying how they can be constructed from a given subsequence using different insertion operations, which are in turn based on analyzing how runs of \texttt{1}'s and \texttt{0}'s can be extended or split. The methods used in the analysis of the underlying combinatorial problems, based on clustering techniques and the run-length encoding of strings may be of interest in their own right. It is thus our hope that our results will also be of independent interest for analyzing estimation and coding problems involving deletion channels.

\subsection{Structure}

We begin by providing a survey of related work in \Cref{sec:related-work}. In \Cref{sec:framework}, we introduce our notation and describe the main definitions, models, and building blocks used in our study. Next, in \Cref{sec:counting-supersequences}, we present an algorithm for counting the number of subsequence embeddings, which relies on the run-length encoding of strings. We then explore counting problems and clustering techniques in the space of supersequences including an analysis of a class of supersequences, referred to as singletons, that admit exactly a single embedding of a given subsequence and prove similar extremization results for their count. We then turn to the original entropy question in \Cref{sec:entropy-minimization} and prove the minimal entropy conjecture for the special cases of single and double deletions. Finally, we conclude by summarizing our findings and stating some open problems in \Cref{sec:conclusion}.

\section{Related Work}\label{sec:related-work}

Studies involving subsequences and supersequences encompass a wide variety of problems that arise in various contexts such as formal languages, coding theory, computer intrusion detection and DNA sequencing to name a few. Despite their prevalence in such a wide range of disciplines, they remain largely unexplored and still present a considerable wealth of unanswered questions.
In the realm of stringology and formal languages, the problem of determining the number of distinct subsequences obtainable from a fixed number of deletions, and closely related problems, have been studied extensively in  \cite{chase1976subsequence,flaxman2004strings,hirschberg1999bounds,hirschberg2000tight}. Perhaps it is worth noting that the same entropy minimizing and maximizing strings conjectured in \cite{atashpendar2015information} and studied in the present work, have been shown to lead to the minimum and maximum number of distinct subsequences, respectively. The problems of finding shortest common supersequences (SCS) and longest common subsequences (LCS) represent two well-known NP-hard problems \cite{jiang1995approximation,middendorf1995finding,middendorf2004combined} that involve similar subproblems. Finally, devising efficient algorithms for subsequence combinatorics based on dynamic programming for counting the number of occurrences of a subsequence in DNA sequencing is yet another important and closely related line of research \cite{rahmann2006subsequence,elzinga2008algorithms}.

In coding theory, and more specifically in the context of insertion and deletions channels, similar long-standing problems have been studied extensively, and yet many problems still remain elusive. This includes designing optimal coding schemes and determining the capacity of deletion channels, both of which incorporate the same underlying combinatorial problem addressed in the present work. The studies in \cite{ullman1967capabilities,swart2003note,kanoria2013optimal} consider a finite number of insertions and deletions for designing correcting codes for synchronization errors and Graham \cite{graham2015binary} studies the problem of reconstructing the original string from a fixed subsequence. More recent works on the characterization of the number of subsequences obtained via the deletion channel can be found in \cite{sala2013counting,sala2015three,cullina2014improvement,liron2015characterization}. Another important body of research in this area is dedicated to deriving tight bounds on the capacity of deletion channels \cite{diggavi2007capacity,kalai2010tight,rahmati2013bounds,cullina2014improvement} and developing bounding techniques \cite{ordentlich2014bounding}.

In terms of more directly related combinatorial objects, Cullina, Kiyavash and Kulkarni \cite{cullina2012coloring} provide a graph-theoretic approach for deletion correcting codes, which among other things, extends Levenshtein's \cite{levenshtein1974elements} result on the size of $\Upsilon_{n,x}$ being only a function of $n$ and $m$, to supersequences of a particular length and Hamming weight. In another more recent work by the same authors \cite{cullina2016restricted}, this result for binary strings is extended to $q$-ary strings of a particular composition, where the composition of a $q$-ary string $x$ refers to a vector of $q$ nonnegative integers, which denote the number of times each symbol in the alphabet appears in the string. The authors \cite{cullina2016restricted} show that the number of distinct supersequences of a particular composition depends only on the composition of the original string, from which the distinct supersequences can be obtained via $n-m$ insertions\footnote{Our independently derived Hamming weight clustering results in Theorem \ref{theorem:maximal-initials} and Theorem \ref{theorem:hamming-cluster} partially overlap with some prior results in \cite{cullina2012coloring} and \cite{cullina2016restricted} by D. Cullina, N. Kiyavash and A. A. Kulkarni.}.

Perhaps rather surprisingly, the problem of determining the number of occurrences of a fixed subsequence in random sequences has not received the same amount and level of attention from the various communities. The state-of-the-art in the finite-length regime remains rather limited in scope. More precisely, the distribution of the number of occurrences constitutes a central problem in coding theory, with a maximum likelihood decoding argument, which represents the holy grail in the study of deletion channels. A comprehensive survey, which among other things, outlines the significance of figuring out this particular distribution is given by Mitzenmacher in \cite{mitzenmacher2009survey}.

\section{Framework}\label{sec:framework}

We consider a memoryless source that emits symbols of the supersequence, drawn independently from the binary alphabet $\Sigma = \{0, 1\}$. Given an alphabet $\Sigma=\{0,1\}$, $\Sigma^n$ denotes the set of all $\Sigma$-strings of length $n$. Let $p_\alpha$ denote the probability of the symbol $\alpha \in \Sigma$ being emitted, which in the binary case simplifies to $p_\alpha = 0.5$. This means that the probability of occurrence of a random supersequence $y$ is given by $P(y) = \prod_{i=1}^n p_{y_i}$. The probability of a subsequence of length $m$ is defined in a similar manner. Throughout, we use $h(s)$ to denote the Hamming weight of the binary string $s$. Throughout, we use the combinatorics of counting multisets to enumerate all possibilities for placing $n$ indistinguishable objects into bins marked by $m$ distinguishable separators such that the resulting configurations are distinguished only by the number of objects present in each bin, which is given by $\binom{n+m-1}{n}$.

\paragraph{Notation} We use the notation $[n] = \{1, 2, \dotsc, n\}$ and $[n_1,n_2]$ to denote the set of integers between $n_1$ and $n_2$; individual bits from a string are indicated by a subscript denoting their position, starting at $1$, i.e., $y = (y_{i})_{i \in [n]} = (y_1, \dotsc, y_n) $. We denote by $|S|$ the size of a set $S$, which for binary strings also corresponds to their length in bits. We also introduce the following notation: when dealing with binary strings, $\alpha^k$ means $k$ consecutive repetitions of $\alpha \in \{0, 1\}$. Throughout, we use $\sigma$ to refer to the constant strings $x=\texttt{1}^m$ and $x=\texttt{0}^m$ for succinctness.

\paragraph{Subsequences and Supersequences} Given $x \in \Sigma^m$ and $y \in \Sigma^n$, let $x = x_1 x_2 \cdots x_m$ denote a subsequence obtained from $y = y_1 y_2 \cdots y_n$ with a set of indexes $1 \le i_1 < i_2 < \cdots < i_m \le n$ such that $y_{i_1} = x_1, y_{i_2} = x_2, \dotsc, y_{i_m} = x_m$. Subsequences are obtained by deleting characters from the original string and thus adjacent characters in a given subsequence are not necessarily adjacent in the original string.

\paragraph{Projection Masks} We define $y_\pi = (y_i)_{i \in \pi} = x$ to mean that the string $y$ filtered by the mask $\pi$ gives the string $x$. Let $\pi$ denote a set of indexes $\{j_1, \dotsc, j_m\}$ of increasing order that when applied to $y$, yields $x$, i.e., $x = y_{j_1} y_{j_2} \cdots y_{j_m}$ and $1 \le j_1 < j_2 \cdots j_m \le n$.

\paragraph{Deletion Masks} A deletion mask $\delta$ represents the set of indexes that are deleted from $y$ to obtain $x$, i.e., $\delta_i \in [n] \setminus \pi$ and $|\delta| = n-m$, whereas a projection mask $\pi$ denotes indexes that are preserved. Thus, similarly, $\delta$ is a subset of $[n]$ and the result of applying a mask $\delta$ on $y$ is denoted by $y_\delta = x$.

\paragraph{Compatible Supersequences} We define the \emph{uncertainty set}, $\Upsilon_{n,x}$, as follows. Given $x$ and $n$, this is the set of $y$ strings that could project to $x$ for some projection mask $\pi$.
\begin{align*}
\Upsilon_{n,x}:=\{y \in \{0,1\}^n: (\exists \pi) [y_\pi=x] \} = \{y \in \{0,1\}^n: (\exists \delta) [y_{\delta}=x] \}
\end{align*}
It was shown by Levenshtein \cite{levenshtein1974elements} that that the cardinality of $\Upsilon_{n,x}$ is independent of the form of $x$ and that it is only a function of $n$ and $m$, i.e.,
\begin{equation}\label{eq:upsilon-cardinality}
|\Upsilon_{n,x}| = \sum_{r=m}^n \binom{n}{r}.
\end{equation}

\paragraph{Number of Masks or Embeddings} Let $\omega_x(y)$ denote the number of distinct ways that $y$ can project to $x$:
\[
\omega_x(y) := | \{ \pi \in \mathcal{P}([n]): y_\pi = x \} |= |\{\delta \in \mathcal{P}([n]): y_{\delta} = x \}|
\]
we refer to the number of masks associated with a pair $(y, x)$ as the weight of $y$, i.e., the number of times $x$ can be embedded in $y$ as a subsequence.

\paragraph{Initial Projection Masks or Canonical Embeddings} Given $y_\pi =x$, we define $\pi$ to be initial if there is no lexicographically earlier mask $\pi'$ such that $y_{\pi'} = x$. $\pi'$ is a lexicographically earlier mask than $\pi$ if, for some $r$, the smallest $r$ members of $\pi$ and $\pi'$ are the same, but the $(r+1)$-th of $\pi'$ is strictly smaller than that of $\pi$. Throughout, we will use $\tilde{\pi}$ to denote an initial projection mask. The first embedding of a subsequence $x$ in $y$ is also often referred to as the \emph{canonical} embedding in the literature. Note that for a fixed mask or embedding $\pi$, the members of $y$ up to the last member of $\pi$ are completely determined if $\pi$ is initial.

\paragraph{Run-Length Encodings} A substring $T$ of a string $Y=y_1 y_2 \ldots y_n$ over $\Sigma$ is called a \emph{run} of $Y$ if $T$ is a consecutive sequence of the same character (i.e., $T \in \alpha^{+}$ for an $\alpha \in \Sigma$). Let $\mathcal{R}_{x, \alpha}$ denote the set of runs of $\alpha$ in $x$. The notion of run-length encoding will be central to our analysis. Given an $n$-bit binary string $y$, its \emph{run-length encoding} (RLE) is the sequence $r_j = (a_j, b_j)$, $1 \leq j \leq m$, such that
\begin{equation*}
y = a_1^{b_1}a_2^{b_2} \cdots a_m^{b_m}, \qquad m \leq n.
\end{equation*}
with $a_j \in \{0,1\}$ and $b_j \in \{1, \dotsc, n\}$. This encoding is unique if we assume that $a_{i} \neq a_{i+1}$, at which point we only need to specify a single $a_i$ (e.g., the first one) to deduce all the others.
Thus the RLE for a string $y$ is denoted by
\begin{equation*}
y = (a_1; b_1, b_2, \dotsc, b_m).
\end{equation*}
When the value of $a_1$ is irrelevant, which will often be the case later on\footnote{Indeed, if $x=y_\pi$, then $\overline{x}=\overline{y}_\pi$ and $\omega_x(y)= \omega_{\overline{x}}(\overline{y})$.}, we will drop it form the notation. Consecutive zeros or ones in a binary string will be referred to as \emph{blocks} or \emph{runs}.
\begin{example}
Let $y = \texttt{0011010001}$; then we have $y = (0; 2, 2, 1, 1, 3, 1)$ as the first bit is zero; and we have 2 zeros, 2 ones, 1 zero, 1 one, 3 zeros, 1 one. Alternatively, $(2, 2, 1, 1, 3, 1)$ designates simultaneously $\texttt{0011010001}$ and $\texttt{1100101110}$.
\end{example}

\paragraph{Entropy} For a fixed subsequence $x$ of length $m$, the underlying weight distribution used in the computation of the entropy is defined as follows. Upon receiving a subsequence $x$, we consider the set of compatible supersequences $y$ of length $n$ (denoted by $\Upsilon_{n,x}$) that can project to $x$ upon $n-m$ deletions. Every $y \in \Upsilon_{n,x}$ is assigned a weight given by its number of masks $\omega_x(y)$, i.e., the number of times $x$ can be embedded in $y$ as a subsequence. We consider the conditional Shannon entropy $H(Y|X=x)$ where $Y$ is confined to the space of compatible supersequences $\Upsilon_{n,x}$. The total number of masks in $\Upsilon_{n,x}$ is given by
\begin{equation}
\label{eq:numbmask}
\mu_{n,m}=\binom{n}{m} \cdot 2^{n-m}
\end{equation}
Let $P_x$ denote the normalized weight distribution given below
\begin{equation*}\label{eq:subseq-prob-distribution1}
P_x=\{P(Y = y | X = x)\text{ for } y \in \Upsilon_{n,x} \}.
\end{equation*}
where $P(Y = y | X = x)$ is given by
\begin{align*}
P(Y = y | X = x) &= \frac{P(Y = y \wedge X = x)}{P(X = x)} = \frac{P(X = x| Y = y) \cdot P(Y = y)}{P(X = x)}
= \frac{\frac{|\{ \pi: \pi(y)=x \}|}{\binom{n}{m}}2^{-n}}{P(X =x)} \\
& = \frac{\omega_x(y)2^{-n}}{\binom{n}{m}P(X =x)} = \frac{\omega_x(y)2^{-n}}{\binom{n}{m}\sum_{y'}P(Y = y')P(X = x | Y = y')} \\
& = \frac{\omega_x(y)2^{-n}}{\binom{n}{m}\sum_{y'}\frac{\omega_x(y')}{\binom{n}{m}}2^{-n}} = \frac{\omega_x(y)}{\sum_{y'}\omega_x(y')} = \frac{\omega_x(y)}{\mu_{n,m}}
\end{align*}
We therefore have \begin{equation}\label{eq:subseq-prob-distribution}
P_x=\Bigg\{\frac{\omega_x(y_1)}{\mu_{n,m}}, \ldots, \frac{\omega_x(y_n)}{\mu_{n,m}} \Bigg\}.
\end{equation}
Finally, for simplicity we use $H_n(x)$ throughout this work to refer to the Shannon entropy of a distribution $P$ corresponding to a subsequence $x$ as defined below
\begin{equation}\label{eq:shannon-entropy}
H_n(x) = -\sum_i p_i \cdot \log_2(p_i)
\end{equation}
where $p_i$ is given by
\begin{equation*}
p_i = \frac{\omega_x(y_i)}{\mu_{n,m}}.
\end{equation*}
An example illustrating these concepts is given in \Cref{table:upsilon-1}. In addition to the distribution of weights, i.e., number of masks per $y$, Hamming-weight groupings of supersequences are indicated by horizontal separators.
\begin{table}[t]
	\tiny
    \caption{Clusters of Supersequences and Distribution of Subsequence Embeddings}
    \label{table:upsilon-1}
	\centering
		\begin{tabular}{|c|c|c|}
		\hline
		 \multicolumn{3}{|c|}{$x=\texttt{110}$}  \\ \hline
		$y$ & $\pi_i$ &$\omega_x(y)$ \\ \hline
		$\texttt{00110}$ & $\{3, 4, 5 \}$ & 1                     \\ \hline
		$\texttt{01010}$ & $\{ 2, 4, 5 \}$ & 1                     \\ \hline
		$\texttt{01100}$ & $\{2, 3, 4\}, \{2, 3, 5\}$ & 2                     \\ \hline
		$\texttt{10010}$ & $\{ 1, 4, 5 \}$ & 1                     \\ \hline
		$\texttt{10100}$ & $\{1, 3, 4\}, \{1, 3, 5\}$ & 2                     \\ \hline
		$\texttt{ 11000 }$ & $\{1, 2, 3\}, \{1, 2, 4\}, \{1, 2, 5\}$ & 3                     \\ \hline \hline \hline
		$\texttt{ 01101 }$ & $\{ 2, 3, 4 \}$  & 1                    \\ \hline
		$\texttt{ 01110 }$ & $\{2, 3, 5\}, \{2, 4, 5\}, \{3, 4, 5\}$ & 3                     \\ \hline
		$\texttt{ 10101 }$ & $\{ 1, 3, 4 \}$ & 1                     \\ \hline
		$\texttt{ 10110 }$ & $\{1, 3, 5\}, \{1, 4, 5\}, \{3, 4, 5\}$ & 3                     \\ \hline
		$\texttt{ 11001 }$ & $\{1, 2, 3\}, \{1, 2, 4\}$ & 2                     \\ \hline
		$\texttt{ 11100 }$ & $\{1, 2, 4\}, \{1, 2, 5\}, \{1, 3, 4\}, \{1, 3, 5\}, \{2, 3, 4\}, \{2, 3, 5\}$ & 6                     \\ \hline \hline \hline
		$\texttt{ 11011 }$ &  $\{ 1, 2, 3 \}$ & 1                    \\ \hline
		$\texttt{ 11101 }$ &  $\{1, 2, 4\}, \{1, 3, 4\}, \{2, 3, 4\}$ & 3                    \\ \hline
		$\texttt{ 11110 }$ &  $\{1, 2, 5 \},\{1, 3, 5 \},\{1, 4, 5 \},\{2, 3, 5 \},\{2, 4, 5 \},\{3, 4, 5 \}$ & 6                    \\ \hline
	\end{tabular}
	\begin{tabular}{|c|c|c|}
		\hline
		\multicolumn{3}{|c|}{$x=\texttt{101}$}  \\ \hline
		$y$ & $\pi_i$ & $\omega_x(y)$  \\ \hline
	  $\texttt{ 00101 }$ & $\{3, 4, 5 \}$ & 1                     \\ \hline
		$\texttt{ 01001 }$ & $\{2, 3, 5\}, \{2, 4, 5 \}$ & 2                     \\ \hline
		$\texttt{ 01010 }$ & $\{ 2, 3, 4 \}$ & 1                     \\ \hline
		$\texttt{ 10001 }$ & $\{1, 2, 5\}, \{1, 3, 5\}, \{1, 4, 5\}$ & 3                     \\ \hline
		$\texttt{ 10010 }$ & $\{1, 2, 4\}, \{1, 3, 4\}$  & 2                    \\ \hline
		$\texttt{ 10100 }$ & $\{ 1, 2, 3 \}$   & 1                   \\ \hline \hline \hline
		$\texttt{ 01011 }$ & $\{ 2, 3, 4 \}, \{ 2,3,5\}$ & 2                     \\ \hline
		$\texttt{ 01101 }$ & $\{ 2, 4, 5 \}, \{3,4,5\}$ & 2                     \\ \hline
		$\texttt{ 10011 }$ & $\{1, 2, 4\}, \{1, 2, 5\}, \{1, 3, 4\}, \{1, 3, 5\}$ & 4                      \\ \hline
		$\texttt{ 10101 }$ & $\{1, 2, 3\}, \{1, 2, 5\}, \{1, 4, 5\}, \{3, 4, 5\}$ & 4                     \\ \hline
		$\texttt{ 11001 }$ & $\{1, 3, 5\}, \{1, 4, 5\}, \{2, 3, 5\}, \{2, 4, 5\}$  & 4                    \\ \hline
		$\texttt{ 11010 }$ & $\{1, 3, 4\}, \{2, 3, 4\}$  & 2                    \\ \hline \hline \hline
		$\texttt{ 10111 }$ &  $\{1, 2, 3\}, \{1, 2, 4\}, \{1, 2, 5\}$ &  3                   \\ \hline
		$\texttt{ 11011 }$ &  $\{1, 3, 4\}, \{1, 3, 5\}, \{2, 3, 4\}, \{2, 3, 5\}$ & 4                    \\ \hline
		$\texttt{ 11101 }$ &  $\{1, 4, 5\}, \{2, 4, 5\}, \{3, 4, 5\}$ & 3                    \\ \hline
	\end{tabular}
\end{table}

\section{Clustering Supersequences, Counting Subsequence Embeddings and Supersequences with Single Embeddings}\label{sec:counting-supersequences}

In the context of the original entropy extremization analysis, the counting problems studied in this section are motivated by the need for gaining a better understanding of the combinatorial objects and structures involving supersequences that exhibit specific properties with respect to a fixed subsequence. Indeed, the quantities of interest in the entropy problem are precisely determined by the number of supersequences that admit a certain embedding weight for a fixed subsequence. Thus, the results in this section are aimed at providing more insight into related combinatoral objects, with similar techniques used in \Cref{sec:entropy-minimization} to cluster supersequences admitting specific weights in order to establish the entropy minimization case.

To this end, we first provide a characterization of the number of subsequence embeddings based on a run-length encoding of strings used for identifying deletion patterns that simplify the counting problem to a sequential mapping of runs from $x$ strings to $y$ strings. We then consider the problem of counting \emph{singletons}, that is, supersequences that admit only a single embedding of $x$. We provide a closed form expression for enumerating singletons using the same run-length encoding and prove an analogous result for the minimization and maximization of the number of singletons, by the alternating and the uniform strings, respectively.

\begin{figure}[tp]
	\centering
	\includegraphics[width=0.8\textwidth]{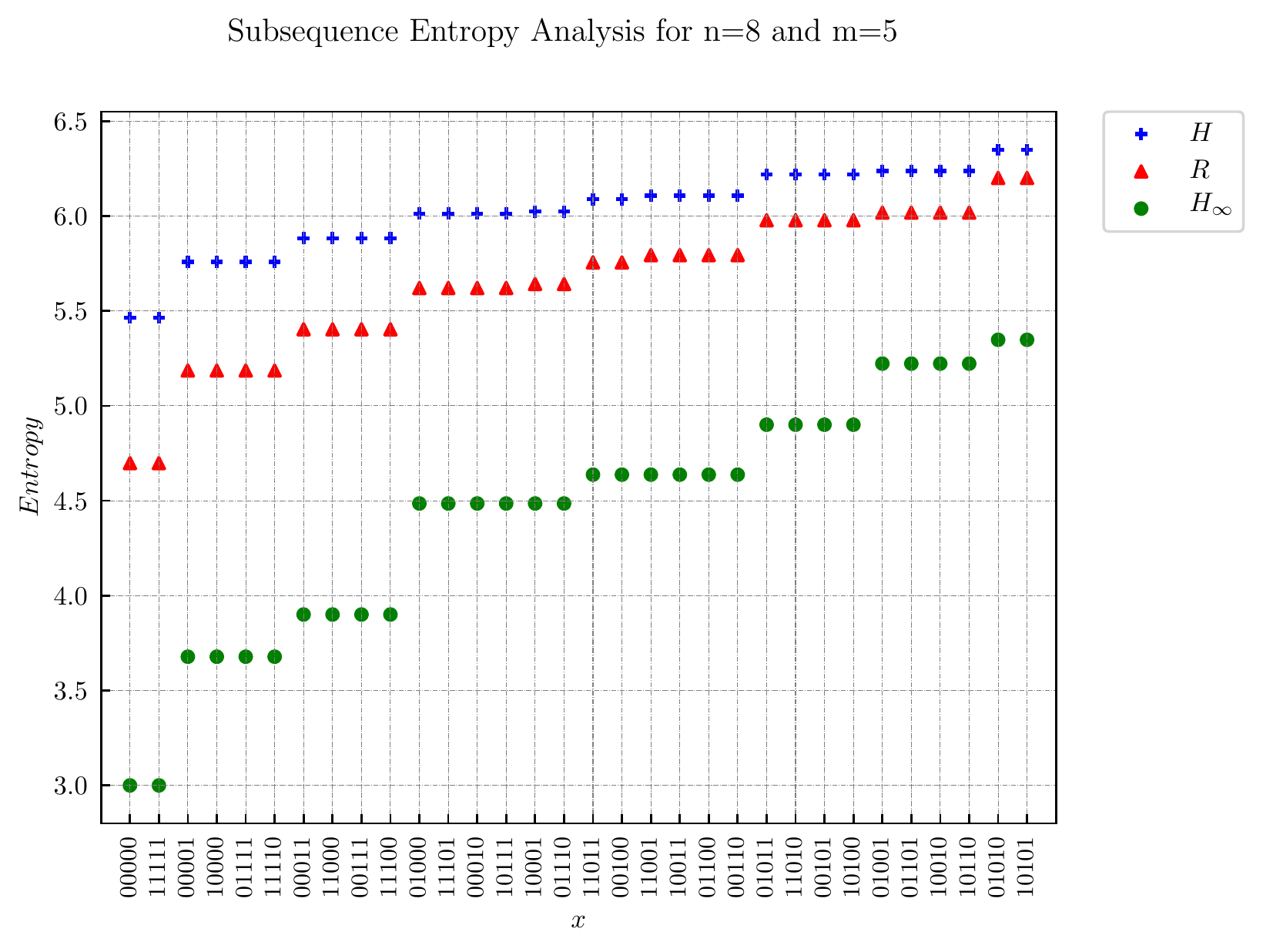}
	\caption{Shannon entropy $H$, second-order R\'{e}nyi entropy $R$, and Min-entropy $H_\infty$}
	\label{figure:entropy}
\end{figure}

We now briefly review the results of the entropy analysis presented in \cite{atashpendar2015information}, in which it is conjectured that the constant/uniform string consisting of all \texttt{1}'s (or all \texttt{0}'s), $x=\texttt{11...1}$, and the alternating $x$ string, i.e., $x=\texttt{1010...}$ minimize and maximize the entropy $H_n(x)$, respectively. To illustrate this, the plot given in \Cref{figure:entropy} shows the values of the min-entropy ($H_\infty$), the second-order R\'{e}nyi entropy ($R$) and the Shannon entropy ($H$) computed for all $x$ strings of length $5$, with $n=8$.

\subsection{Counting Subsequence Embeddings via Runs}\label{sec:subseq-count-formalism}

Efficient dynamic programming algorithms for computing the number of subsequence embeddings are known in the literature, e.g., a recursive algorithm requiring $\Theta(n \times m)$ operations \cite{elzinga2008algorithms}. Here we provide an alternative algorithm, which is primarily based on the run-length encoding of strings.

Using the RLE notation, there are a few cases in which this question is easy to answer. For instance, if $y = (a;k_1, \dotsc, k_\ell)$ and $x = (a;k_1', \dotsc, k_\ell')$, with the same value of $\ell$, i.e., we have the same number of blocks in $x$ and $y$, then it is easy to see that there is a one-to-one sequential mapping of blocks between $x$ and $y$. This allows us to enumerate the different masks depending on how they map the blocks to each other as follows:
\begin{equation}\label{eq:embedding-simple-case}
\omega_x(y) = \prod_{i=1}^\ell \binom{k_i}{k'_i}.
\end{equation}
However, in the general case, the number of blocks in $x$ and $y$ can be different. If $y= (a;k_1, \dotsc, k_\ell)$ and $x= (\overline{a};k'_1, \dotsc, k'_{\ell'})$ do not start with the same character, we have to delete the first block to recover the case $y= (a;k_2, \dotsc, k_{\ell})$, and $x= (a;k'_1, \dotsc, k'_{\ell'})$. We will now suppose that $x$ and $y$ start with the same character.

Here we describe an algorithm wherein for a fixed pair of $x$ and $y$ strings, we structure and enumerate the corresponding space of masks by accounting for the number of different ways we can delete characters in order to merge blocks/runs such that we can recover the simple case given in \Cref{eq:embedding-simple-case}. In the more general case, let $y=(k_1,  \dotsc, k_{\ell})$ and $x=(k'_1, \dotsc, k'_{\ell'})$.

\begin{definition}
Let $S$ be the set of maps $f: [\ell'] \to [\ell]$ that satisfy the following properties: $f$ is strictly increasing and $f(i) \equiv i \bmod 2$. A function $f$ will define a subset of masks, by specifying blocks that will have to be completely deleted. We group the masks according to a set of functions $f$ that map indexes of blocks of $x$ to indexes of blocks of $y$. Intuitively, $f$ maps the $i$-th block of $x$ to the block of $y$ that contains the last character of the $i$-th block of $x$.
Therefore, all blocks of $y$ between $f(i) +1 $ and $f(i)$ that are not composed of matching characters have to be deleted such that we can recover the simple case in \Cref{eq:embedding-simple-case}.
\end{definition}
For the subsequent analysis, recall that $k_i$ denotes the length of the run at index $i$, whereas $k^*_i$ refers to the actual set of indexes of the $i$-th run.
\begin{definition}
Let $k^*_{i}$ denote the set of indexes belonging to the $i$-th block of $y$, i.e., $\{ \sum_{j=1}^{i-1} k_j, \sum_{j=1}^{i-1} k_j + 1, \dotsc, \sum_{j=1}^{i} k_j \}$, $F^{*}(i) = \{f(i-1) + 2, f(i-1) + 4, \dotsc, f(i) - 1\}$ and $\overline{F^*(i)} = \{f(i-1) + 1, f(i-1) + 3, \dotsc, f(i)\}$, then a deletion mask $\delta$ corresponds to $f$ if:
\begin{itemize}
\item $\forall i: \cup_{j \in F^*(i)} k^*_{j} \subset \delta$
\item $\forall i: k^*_{f(i)} \not\subset \delta$ (this allows us to have a partition)
\end{itemize}
We call $\omega_f$ the set of masks corresponding to $f$.
\end{definition}
\begin{theorem}
The family $(\omega_f)_{f \in S}$ defines a partition on the set of masks from $y$ to $x$.
\end{theorem}
\begin{IEEEproof}
We first show that for $f \neq f' \in S$, every deletion mask $\delta$ corresponding to $f$ is different from every mask $\delta'$ associated with $f'$ (i.e., $\omega_f \cap \omega_{f'} = \emptyset$).
Since $f \neq f'$, we have a \emph{smallest} integer $i \in [\ell]$ such that $f(i) \neq f'(i)$. We assume without loss of generality that $ f'(i-1)=f(i-1)<f(i) < f'(i)$. Due to the condition on parity, $f(i+1) \neq f'(i)$.
We distinguish between two cases:
\begin{itemize}
\item If $f'(i-1)<f(i+1)<f'(i)$, then $k_{f(i+1)} \not\subset \delta$, and $k_{f(i+1)} \subset \delta'$ since $f(i+1) \in F'^*(i-1)$.

\item Conversely if $f(i-1)<f'(i)< f(i+1)$, then $k_{f'(i)} \not\subset \delta'$, and $k_{f'(i)} \subset \delta$ since $f'(i) \in F^*(i+1)$.
\end{itemize}
Therefore, we have $\delta \neq \delta'$. We now show that $\cup_{f \in S}\omega_f$ is the set of masks from $y$ to $x$. We will use projection masks here as they are more suitable for this proof. Let $\pi$ be a projection mask such that $y_{\pi} = x$. We let $\pi = \{\pi_1, \dotsc, \pi_m\}$, where the $\pi_i$ are in increasing order. Therefore, we have for all $i$, $y_{\pi_i} = x_i$. We define $\phi :[n] \rightarrow [\ell]$ to be the mapping that takes an index of $y$ and returns the index of the block/run it belongs to, i.e., $\phi(a)$ returns the smallest $i$ such that $\sum_{j=1}^i k_j \geq a$.
We define $f$ such that $f \in S$ and $\pi$ is in $\omega_f$, by $f(i) = \phi( \pi_{\sum_{j=1}^i k'_j})$. To prove that $f$ is in $S$, note that given $i$:
\begin{itemize}
\item We have that $f(i) \leq f(i+1)$ since the $\pi_i$ are in increasing order.
\item Moreover, $\pi_{\sum_{j=1}^i k'_j}$ and $\pi_{\sum_{j=1}^{i+1} k'_j}$ correspond to indexes (of $y$) of opposite letter (if the first one is a $1$, the second is a $0$ and vice versa) since $\sum_{j=1}^i k'_j$  and $\sum_{j=1}^{i+1} k'_j$ correspond to indexes (of $x$) of opposite letter.
\end{itemize}
Therefore, $f(i)$ and $f(i+1)$ are of opposite parity and $f(i) < f(i+1)$.

We now prove that $\pi$ corresponds to $f$. For a fixed $i \in [\ell]$, let $k^*_{f(i-1)} = b^{k_{f(i-1)}}$, i.e., the $f(i-1)$-th block of $y$ is made of letters $b$. Therefore, $k^*_{t} = b^{k_t}$ for $t \in F^*(i)$, since $t$ has the same parity as $f(i-1)$. Moreover, we have $b = x_{\sum_{j=1}^{i-1} k'_j}$ according to the definition of $f$. So for every index $h$ between $\sum_{j=1}^{i-1} k_j +1$ and $\sum_{j=1}^{i} k_j$, $x_h = y_{\pi_h} = \overline{b}$, and for $t \in F^*(i)$, we have $k_t^* \cap \pi = \emptyset$ (equivalently with the deletion mask $\delta$, $k_t^* \subset \delta$). By definition, $\pi_{\sum_{j=1}^{i} k_j} \in k_{f(i)}^*$ so $\pi \cap k_{f(i)}^* \neq \emptyset$ (equivalently with the deletion mask $\delta$, $k_{f(i)}^* \not\subset \delta$).
\end{IEEEproof}
\begin{theorem}
We have for $f\in S$, given by
\begin{equation}
	|\omega_f| = \prod_{i=0}^{\ell} \binom{\sum_{j \in \overline{F^*(i)}}k_j }{k'_i} - \binom{\sum_{j \in \overline{F^*(i)}\setminus \{f(i)\}}k_j }{k'_i}
\end{equation}

\end{theorem}
\begin{IEEEproof}
Upon the deletion induced by $f$, we obtain a string of the form $(\sum_{j \in \overline{F^*(1)}}k_j, \dotsc , \sum_{j \in \overline{F^*(\ell)}}k_j)$. Therefore, we have the same number of blocks in both the $y$ string as well as the $x$ string, and the number of masks can be computed easily as shown in \Cref{eq:embedding-simple-case}. We first count the number of ways to choose $k_i'$ elements from $k_{\overline{F^*(i)}}$ and then subtract the number of combinations not using any of the $k_{f(i)}$.
\end{IEEEproof}
\begin{remark}
We can note that $\overline{F^*(i)}$ and $F^*(i)$ form a partition of $[\ell]$.
\end{remark}
Following from the preceding theorems, the total number of masks can be computed as follows
\begin{equation}
	\omega_x(y) = \sum_{f \in S} |\omega_f|.
\end{equation}
By summing over all $f \in S$, we get the total number $|\omega|$ of compatible masks. Note that it may happen that $|\omega_f| = 0$; this happens when we try to trace a large block of $x$ from a smaller block of $y$.

\textbf{The Set $S$:} We now determine the size of $S$, as a function of $\ell$ and $\ell'$. Let this size be denoted by $\sigma(\ell',\ell)$. We denote $u = \lfloor (\ell-\ell')/2 \rfloor$.
If $f(1) = 1$, then we get $\sigma(\ell'-1,\ell-1)$; if $f(1) =3$, we get $\sigma(\ell'-1,\ell-3)$, etc. We also know that $\sigma(x,x) = 1$ for all $x$, and that $\sigma(x,y) = 0$ for all $x$, $y$ such that $y<x$.
We therefore get the following recurrence:
\begin{equation*}
	\sigma(\ell',\ell) = \sum_{i=0}^{u} \sigma(\ell'-1,\ell-1-2i)
\end{equation*}
Iterating this recursion, we get
\begin{equation*}
\sigma(\ell',\ell) = \sum_{i=0}^{u}\sum_{j=0}^{u -i} \sigma(\ell'-2,\ell-2-2i-2j),
\end{equation*}
and grouping the terms yields
\begin{equation*}
	\sigma(\ell',\ell) = \sum_{i=0}^{u} (i+1)\sigma(\ell'-2,\ell-2-2i).
\end{equation*}
We now describe a direct combinatorial argument which gives a closed form formula for $\sigma(\ell',\ell) = |S|$. First note that if $\ell\not\equiv \ell'$ mod 2 then $\ell$ cannot be in the image of $f$. So let $\tilde{\ell}=\ell$ if $\ell\equiv \ell'$ mod 2 and $\tilde{\ell}=\ell-1$ if not. Now the problem is to choose $[\ell']$ elements from $[\tilde{\ell}]$ such that all the gaps have even width. Equivalently, we are interleaving the $\ell'$ chosen elements with $u=(\tilde{\ell}-\ell')/2$ gap-segments of width 2. The number of ways to do this is plainly
\begin{equation}
|S|=\sigma(\ell', \ell)=\binom{\ell'+u}{u}.
\end{equation}

\begin{example}
For	$y=	\texttt{0000111100001111}$ and $x=\texttt{0011}$, we obtain $\omega_y(x)=300$. We now compute the number of embeddings using the run-based algorithm described above. We have $\ell=4$, $\ell'=2$ and $u=(\tilde{\ell}-\ell')/2$, which means the size of $S$ is $|S|=\sigma(\ell',\ell)=\binom{l'+u}{u}=\binom{2+1}{1}=3$. The three deletions $S=\{f_1, f_2, f_3\}$ are computed as follows: $y_{f_1} = (k_1, k_2) = \texttt{00001111}$, which amounts to $\omega_{f_1}=\binom{4}{2}\binom{4}{2}=36$. Similarly, for $f_2$ and $f_3$, we get $y_{f_1} = (k_1 + k_3, k_4) = \texttt{000000001111}$ and $y_{f_1} = (k_1, k_2 + k_4) = \texttt{000011111111}$, the two of which add up to $2\times\big(\binom{8}{2}\binom{4}{2}-\binom{4}{2} \big) = 2 \times 132 = 264$. So the total is $\omega_y(x) = \sum_{f \in S} \Omega_f = 36+132+132=300$.
\end{example}

\subsection{From Maximal Initials to Hamming Clusters}
\begin{definition}
Let $\Upsilon_{n,x}^c$ be the cluster of supersequences that have $c$ extra \texttt{1}'s with respect to $x$, where $0 \le c \le n-m$.
\[
\Upsilon^c_{n,x} = \{ y \in \Upsilon_{n,x} \mid h(y) - h(x) = c \}.
\]
\end{definition}
The set of compatible supersequences is thus broken down into $n-m+1$ disjoint sets indexed from 0 to $n-m$ such that strings in cluster $c$ contain $h(x)+c$ \texttt{1}'s:
\begin{equation*}
\Upsilon_{n,x} = \bigcup\limits_{c=0}^{n-m} \Upsilon^c_{n,x}.
\end{equation*}
\begin{definition}
Maximal initials represent $y$ strings for which the largest index of their initial mask, $\tilde{\pi}$, overlaps with the last bit of $y$. In other words, the last index of the canonical embedding of $x$ in $y$ overlaps with the last bit of $y$. Recall that we use $\tilde{\pi}$ to denote a mask $\pi$ that is initial.
\[
\mathcal{M}_{n,x} = \left\{ y \in \Upsilon_{n,x} \mid (\exists \tilde{\pi})[ y_{\tilde{\pi}}=x \wedge \max(\tilde{\pi}) = |y| = n ] \right\}.
\]
Similarly, we define a clustering for maximal initials based on the Hamming weight of the $y$ strings
\[
\mathcal{M}_{n,x}^c = \{ y \in \mathcal{M}_{n,x} \mid h(y) = h(x)+c \}.
\]
\end{definition}
\begin{example}
For example, the initial embedding of  $x=\texttt{1011}$ in $y=\texttt{110011}$ given by $\tilde{\pi}=\{1,3,5,6\}$ is maximal, whereas its initial embedding in $y'=\texttt{101011}$ given by $\tilde{\pi}'=\{1,2,3,5\}$ is not maximal as the last index of $\tilde{\pi}'$ does not overlap with the position of the last bit of $y'$.
\end{example}
A more exhaustive example illustrating these concepts is given in \Cref{table:upsilon-2}. In addition to the distribution of weights, i.e., number of masks per $y$, clusters and maximal initials are indicated by horizontal separators and bold font, respectively.
\begin{table}[t]
    \caption{Clusters, maximal initial projection masks and distribution of weights.}
    \label{table:upsilon-2}
	\centering
		\begin{tabular}{|c|c|c|}
		\hline
		 \multicolumn{3}{|c|}{$x=\texttt{110}$}  \\ \hline
		$y$ & $\tilde{\pi}$ &$\omega$ \\ \hline
		$\texttt{00110}$ & $\mathbf{\{3, 4, 5 \}}$ & 1                     \\ \hline
		$\texttt{01010}$ & $\mathbf{\{ 2, 4, 5 \}}$ & 1                     \\ \hline
		$\texttt{01100}$ & $\{ 2, 3, 4 \}$ & 2                     \\ \hline
		$\texttt{10010}$ & $\mathbf{\{ 1, 4, 5 \}}$ & 1                     \\ \hline
		$\texttt{10100}$ & $\{ 1, 3, 4 \}$ & 2                     \\ \hline
		$\texttt{ 11000 }$ & $\{ 1, 2, 3 \}$ & 3                     \\ \hline \hline \hline
		$\texttt{ 01101 }$ & $\{ 2, 3, 4 \}$  & 1                    \\ \hline
		$\texttt{ 01110 }$ & $\mathbf{\{ 2, 3, 5 \}}$ & 3                     \\ \hline
		$\texttt{ 10101 }$ & $\{ 1, 3, 4 \}$ & 1                     \\ \hline
		$\texttt{ 10110 }$ & $\mathbf{\{ 1, 3, 5\}}$ & 3                     \\ \hline
		$\texttt{ 11001 }$ & $\{ 1, 2, 3 \}$ & 2                     \\ \hline
		$\texttt{ 11100 }$ & $\{ 1, 2, 4\}$ & 6                     \\ \hline \hline \hline
		$\texttt{ 11011 }$ &  $\{ 1, 2, 3 \}$ & 1                    \\ \hline
		$\texttt{ 11101 }$ &  $\{ 1, 2, 4 \}$ & 3                    \\ \hline
		$\texttt{ 11110 }$ &  $\mathbf{\{1, 2, 5 \}}$ & 6                    \\ \hline
	\end{tabular}
	\begin{tabular}{|c|c|c|}
		\hline
		\multicolumn{3}{|c|}{$x=\texttt{101}$}  \\ \hline
		$y$ & $\tilde{\pi}$ & $\omega$  \\ \hline
		$\texttt{ 00101 }$ & $\mathbf{\{3, 4, 5 \}}$ & 1                     \\ \hline
		$\texttt{ 01001 }$ & $\mathbf{\{ 2, 3, 5 \}}$ & 2                     \\ \hline
		$\texttt{ 01010 }$ & $\{ 2, 3, 4 \}$ & 1                     \\ \hline
		$\texttt{ 10001 }$ & $\mathbf{\{ 1, 2, 5 \}}$ & 3                     \\ \hline
		$\texttt{ 10010 }$ & $\{ 1, 2, 4 \}$  & 2                    \\ \hline
		$\texttt{ 10100 }$ & $\{ 1, 2, 3 \}$   & 1                   \\ \hline \hline \hline
		$\texttt{ 01011 }$ & $\{ 2, 3, 4 \}$ & 2                     \\ \hline
		$\texttt{ 01101 }$ & $\mathbf{\{ 2, 4, 5 \}}$ & 2                     \\ \hline
		$\texttt{ 10011 }$ & $\{ 1, 2, 4 \}$ & 4                      \\ \hline
		$\texttt{ 10101 }$ & $\{ 1, 2, 3\}$ & 4                     \\ \hline
		$\texttt{ 11001 }$ & $\mathbf{\{ 1, 3, 5 \}}$  & 4                    \\ \hline
		$\texttt{ 11010 }$ & $\{ 1, 3, 4\}$  & 2                    \\ \hline \hline \hline
		$\texttt{ 10111 }$ &  $\{ 1, 2, 3 \}$ &  3                   \\ \hline
		$\texttt{ 11011 }$ &  $\{ 1, 3, 4 \}$ & 4                    \\ \hline
		$\texttt{ 11101 }$ &  $\mathbf{\{1, 4, 5 \}}$ & 3                    \\ \hline
	\end{tabular}
\end{table}
\begin{theorem}\label{theorem:total-maximal-initials}
For given $n$, the cardinality of $\mathcal{M}_{n,x}$ is independent of the exact $x$.
\end{theorem}
\begin{IEEEproof}
It is clear that every $n$-element sequence that has $x$ as an $m$-element subsequence has a unique initial mask $\tilde{\pi}$ that gives $x$. Furthermore, if we fix $\pi$, then the members of $y$ up to the last member of $\pi$ are completely determined if $\pi$ is initial. To see this, consider the case $i \in \tilde{\pi}$, then $y_i$ (the $i$-th member of $y$) must correspond to $x_j$, where $i$ is the $j$-th smallest member of $\tilde{\pi}$. If $i \notin \tilde{\pi}$, but smaller than $max(\tilde{\pi})$, then the $i$-th member of $y$ must correspond to $x_{j+1}$, where $j$ is the number of members of $\tilde{\pi}$ smaller than $i$. The latter follows because if this bit were $x_{j+1}$, then the given $\pi$ would not be initial.

We also need to observe that for a given $\tilde{\pi}$, there always exists a $y$ that has $x$ initially in $\tilde{\pi}$: suppose that $x$ starts with a \texttt{0}, we set all the bits of $y$ before $\tilde{\pi}$ to be \texttt{1}. For a given value $\ell$ of $max(\tilde{\pi})$ - which can range from $m$ to $n$ - there are exactly $\binom{\ell-1}{m-1}$ $\tilde{\pi}$'s, one for each selection of the other $m-1$ members of $\tilde{\pi}$ amongst the $\ell-1$ values less than $\ell$.

Moreover, here we have an additional constraint, namely that the initial masks should be maximal as well, i.e., $max(\tilde{\pi}) = n$. This means that $\ell = n$ and so we can count the number of distinct initials for the remaining $m-1$ elements of $x$ in the remaining $(n-1)$-long elements of $y$ strings, which is simply given by
\begin{equation}
|\mathcal{M}_{n,x}|=|\mathcal{M}_{n,m}|=\binom{n-1}{m-1}
\end{equation}
Clearly the cardinality of the set of maximal initials is independent of the form of $x$ and depends only on $n$ and $m$.
\end{IEEEproof}

\begin{remark}
Note that if we extend the analysis in the proof of \Cref{theorem:total-maximal-initials} and let $\ell$ run over the range $[m,n]$, we can count all the distinct initial embeddings in $\Upsilon_{n,x}$, given by $\sum_{\ell=m}^n\binom{\ell-1}{m-1}$.

Moreover, since the bits beyond $max(\tilde{\pi})$ are completely undetermined, for a given $\tilde{\pi}$, there are exactly $2^{n-max(\tilde{\pi})}$ $y$'s that have $\tilde{\pi}$ in common, which, incidentally, provides yet another proof for the fact that $|\Upsilon_{n,x}|$ is a function of only $n$ and $m$ since $|\Upsilon_{n,x}|= \sum_{\ell=m}^n\binom{\ell-1}{m-1}2^{n-\ell}$. This allows us to choose the $x$ comprising $m$ \texttt{0}'s and the result in \Cref{eq:upsilon-cardinality} follows immediately.
\end{remark}

\begin{theorem}\label{theorem:maximal-initials}
All $x$ strings of length $m$ that have the same Hamming weight, give rise to the same number of maximal initials in each cluster.
\[
\forall x, x' \in \Sigma^m, h(x) = h(x') \implies |\mathcal{M}^c_{n,x}| = |\mathcal{M}^c_{n, x'}|.
\]
\end{theorem}
\begin{IEEEproof}
We now describe a simple combinatorial argument for counting the number of maximal initials in each cluster indexed by $c$, i.e., a grouping of all $y \in \Upsilon^c_{n,x}$ such that $h(y) = h(x)+c$. Let $p$ and $q$ denote the number of additional \texttt{0}'s and \texttt{1}'s contributed by each cluster, respectively. Furthermore, let $a$ and $b$ denote the number of $\texttt{1}$'s and $\texttt{0}$'s in $x$, respectively.

Similar to the method used in the proof of Theorem \ref{theorem:total-maximal-initials}, due to maximality we fix the last bit of $y$ and $x$, and consider $y'=y-tail(y)$ and $x'=x-tail(x)$ where $tail(s)$ denotes the last bit of $s$. Now the problem amounts to counting distinct initials of length $m-1$ in $(n-1)$-long elements in each cluster by counting the number of ways distinct configurations can be formed as a result of distributing $c$ \texttt{1}'s and $(n-m-c)$ \texttt{0}'s around the bars/separators formed by the $b$ $\texttt{0}$'s and $a$ $\texttt{1}$'s in $x$, respectively.

We now need to observe that to count such strings with distinct initials, we can fix the $m-1$ elements of $x'$ as distinguished elements and count all the unique configurations formed by distributing $p$ indistinguishable $\texttt{0}$'s and $q$ indistinguishable $\texttt{1}$'s among bins formed by the fixed \texttt{1}'s and \texttt{0}'s of $x'$ such that each such configuration is distinguished by a unique initial.

Equivalently, we are counting the number of ways we can place the members of $x'$ among $n-1$ positions comprising $p$ \texttt{0}'s and $q$ \texttt{1}'s without changing the relative order of the elements of $x'$ such that these configurations are uniquely distinguished by the positions of the $m-1$ elements.

Intuitively, the arrangements are determined by choosing the positions of the $m-1$ bits of $x'$: by counting all the unique distributions of bits of opposite value around the elements of $x'$, we are simply displacing the elements of $x'$ in the $n-1$ positions, thereby ensuring that each configuration corresponds to a unique initial.

Note that this coincides exactly with the multiset coefficient (computed via the method of stars and bars) as we can consider the elements of the runs of $x$ to be distinguished elements forming bins among which we can distribute indistinguishable bits of opposite value to count the number of configurations that are distinguished only by the number of $\texttt{1}$'s and $\texttt{0}$'s present in the said bins.

Thus we count the number of unique configurations formed by distributing $p$ \texttt{0}'s and $q$ \texttt{1}'s among the $a$ \texttt{1}'s and $b$ \texttt{0}'s of $x$, respectively. The total count for each cluster $c$ is given by: $\binom{p+a-1}{p} \binom{q+b-1}{q}$, which expressed in terms of the Hamming weight of $x$ gives
\begin{equation}\label{eq:exp-maximal-initial}
|\mathcal{M}^c_{n,x}| = \binom{(n-m-c)+h(x)-1}{n-m-c} \binom{c+(m-h(x))-1}{c}
\end{equation}
With the total number of maximal initials in $\Upsilon_{n,x}$ given by
\[
|\mathcal{M}_{n,x}| = \sum_{c=0}^{n-m} |\mathcal{M}^c_{n,x}|=\binom{n-1}{m-1}.
\]
\end{IEEEproof}

\begin{theorem}\label{theorem:hamming-cluster}
The size of a cluster is purely a function of $n, m, c$ and $h(x)$
\[
\forall x, x' \in \Sigma^m, h(x) = h(x') \implies |\Upsilon^c_{n,x}| = |\Upsilon^c_{n, x'}|
\]
\end{theorem}
\begin{IEEEproof}
 Let $\ell$ denote the position of the last bit of $y$ ranging from $|x|=m$ to $|y|=n$. Starting from a fixed $x$ string, we enumerate all $y$ strings in cluster $c$ by considering maximal initials within the range of $\ell$, i.e., $\ell \in [m, \ldots, n]$.

 Let $g$ denote the number of \texttt{1}'s belonging to the surplus bits in cluster $c$ constrained within the range of the maximal initial, $[1, \dotsc, \ell]$. For each $\ell$, compute $|\mathcal{M}^g_{\ell,x}|$ and count the combinations of choosing the remaining $c-g$ additional bits in the remaining $n-\ell$ bits. Let $UB = \min(c, \ell-m)$ and $LB = \max(0, c-(n-\ell))$ and thus we get the following:
\begin{equation}\label{eq:exp-cluster-raw}
|\Upsilon^c_{n,x}| = \sum_{\ell=m}^{n} \sum_{g=\max\left(0, c-\left(n-\ell\right)\right)}^{\min(c, \ell-m)} |\mathcal{M}^g_{\ell,m}|
\binom{n-\ell}{c-g}
\end{equation}
Finally, inserting \Cref{eq:exp-maximal-initial} into \Cref{eq:exp-cluster-raw} gives
\begin{equation}\label{eq:exp-cluster}
|\Upsilon^c_{n,x}| = \sum_{\ell=m}^{n} \sum_{g=LB}^{UB} \binom{(\ell-m-g)+h(x)-1}{\ell-m-g}
\binom{g+(m-h(x))-1}{g}
\binom{n-\ell}{c-g}.
\end{equation}
As shown in \Cref{eq:exp-cluster}, $|\Upsilon^c_{n,x}|$ depends on the length and the Hamming weight of $x$, but it is independent of the exact form of $x$.
\end{IEEEproof}

\subsection{Simple closed form expression for the size of a cluster}
We have shown that $|\Upsilon^c_{n,x}|$ is independent of the form of $x$. We can now derive a more simplified analytic expression for this count by considering an $x$ string of the following form: $x =\mathtt{11...11}_a\mathtt{00...0}_{m}$, i.e., $a$ \texttt{1}'s followed by $b$ \texttt{0}'s, with $a > 0$ and $b=m-a$.

The $y$ strings in each cluster are precisely the strings of length $n$ that have $a+c$ \texttt{1}'s in them (and $n-a-c$ \texttt{0}'s) where the $a$-th \texttt{1} (i.e., the last one in an initial choice for $x$) occurs before at least $b$ \texttt{0}'s. Clearly there are $\binom{n}{a+c}$ strings with exactly $a+c$ \texttt{1}'s, but some of these will violate the second principle. To find an expression for counting the valid instances, we sum over the positions of the $a$-th \texttt{1}, which must be between $a$ and $a+z$, where $z = n-a-b-c$ is the number of added \texttt{0}'s. Thus we get the following expression
\begin{equation}
|\Upsilon^c_{n,x}|=\sum_{p=h(x)}^{h(x)+z} \binom{p-1}{h(x)-1} \binom{n-p}{c}.
\end{equation}
With $z=n-m-c$ and $p$ denoting the index of the $a$-th 1, we thus count the number of ways of picking \texttt{1}'s before $p$ and the $c$ \texttt{1}'s after $p$. Note that for $h(x)=0$, the cardinality of cluster $c$ is simply given by $\binom{n}{c}$.

\subsection{Recursive expression for the size of a cluster}
We present a recurrence for computing the size of a cluster by considering overlaps between the first bits of $x$ and $y$, respectively. Let $\bullet$ and $\varepsilon$ denote concatenation and the empty string, respectively. Moreover, let $x'$ be the tail of $x$ (resp. $y'$ the tail of $y$).
\begin{itemize}
\item $\Upsilon^c_{n,\texttt{0}\bullet x} = \Upsilon^c_{n-1,x} + \Upsilon^{c-1}_{n-1,\texttt{0}\bullet x}$
\begin{itemize}
\item First term: first bit of $y$ is 0, find $x'$ in $y'$
\item Second term: first bit of $y$ is 1 (part of cluster), so we reduce $c$ and find $x$ in $y'$
\end{itemize}
\item $\Upsilon^c_{n,\texttt{1}\bullet x} = \Upsilon^c_{n-1,x} + \Upsilon^{c}_{n-1,\texttt{1}\bullet x}$
\begin{itemize}
\item Same arguments as above, but for $x$ starting with 1
\end{itemize}
\item Base cases:
\begin{itemize}
\item $\Upsilon^0_{n,\texttt{0}\bullet x} = \Upsilon^0_{n-1,x}$
\item $\Upsilon^0_{n,\texttt{1}\bullet x} = \Upsilon^0_{n-1,x} + \Upsilon^{0}_{n-1,\texttt{1}\bullet x}$
\item $\Upsilon^c_{n,\varepsilon} = \binom{n}{c}$
\item if $c+|x| > n$ then return 0 else $\Upsilon^c_{n,x}$
\end{itemize}
\end{itemize}
It is worth pointing out that since this recursion depends on the form of $x$, i.e., whether or not $x$ starts with a 0 or 1, it does not explicitly capture the bijection between clusters of $x$ strings that have the same Hamming weight, as proved in \Cref{theorem:hamming-cluster}.

\subsection{Enumerating Singletons via Runs}\label{sec:singletons}

Let \emph{singletons} define supersequences in $\Upsilon_{n,x}$ that admit exactly a single mask for a fixed subsequence $x$ of length $m$, i.e., they give rise to exactly a single occurrence of $x$ upon $n-m$ deletions. We use $\mathcal{S}_{n,x}$ to denote this set.
\begin{equation*}
\mathcal{S}_{n,x} = \{ y \in \Upsilon_{n,x} | \; \omega_x(y) = 1 \}.
\end{equation*}
To compute the cardinality of $\mathcal{S}_{n,x}$, we describe a counting technique based on splitting runs of \texttt{1}'s and \texttt{0}'s in $x$ according to the following observations: $(i)$ inserting bits of opposite value to either side of the framing bits in $x$, i.e., before the first or after the last bit of $x$, does not alter the number of masks. $(ii)$ splitting runs of $0$'s and $1$'s in $x$, i.e., insertion of bits of opposite value in between two identical bits, does not modify the count. This amounts to counting the number of ways that singletons can be obtained from a fixed $x$ string via run-splitting insertions.

The number of possible run splittings corresponds to the number of distinct ways that $c$ \texttt{1}'s and $(n-m-c)$ \texttt{0}'s can be placed in between the bits of the runs of \texttt{0}'s and \texttt{1}'s in $x$, respectively. Again, this count is given by the multiset number $\binom{a+b-1}{a}$, where we count the number of ways $a$ indistinguishable objects can be placed into $b$ distinguishable bins. Note that the number of singletons depends heavily on the number of runs in $x$ and their corresponding lengths. The counting is done by summing over all $n-m$ cases and computing the configurations that lead to singletons as a function of the runs in $x$ and the number of additional \texttt{1}'s and \texttt{0}'s that can be inserted into $x$.

In order to do this computation, we first count the number of insertions slots in $x$ as a function of its runs of \texttt{1}'s and \texttt{0}'s, given by $\rho_0(x)$ and $\rho_1(x)$, respectively. Let $r_i^j$ be a run with $i$ and $j$ denoting its first and last index and let $\rho_\alpha(x)$ denote the number of insertion slots in $x$ as a function of its runs of $\alpha$. To compute $\rho_\alpha(x)$, we iterate through the runs of $\alpha$ and in $x$ and count the number of indexes at which we can split runs as follows
\begin{align}
\rho_\alpha(x) & = \sum_{r \in \mathcal{R}_{x,\alpha}} f(r)
\end{align}
where
\begin{equation}
f(r)=
\begin{cases}
|r_i^j|+1,& \text{if } i=1 \wedge j=n\\
|r_i^j|,& \text{if } (i=1 \wedge j<n) \vee (i>1 \wedge j=n)\\
|r_i^j| - 1,              & \text{otherwise}
\end{cases}
\end{equation}
Note that if either the first bit or the last bit of a run overlaps with the first or last bit of $x$, the number of bars is equal to the length of the run. If the said indexes do not overlap with neither the first nor the last bit of $x$, the count is equal to the length of the run minus 1, and finally if both indexes overlap with the first and last bit of $x$ the count is equal to the length of the run plus 1.

We can now count the total number of singletons for given $n$ and $x$ as follows. Let $c$ and $b$ ($b=n-m-c$) denote the number of \texttt{1}'s and \texttt{0}'s contributing to the insertions, and the total number of singletons is given by
\begin{equation}
|\mathcal{S}_{n,x}| = \binom{n-m + \rho_1(x) - 1}{n-m} + \sum_{c=1}^{n-m-1} \binom{b+\rho_1(x)-1}{b} \binom{c+\rho_0(x)-1}{c} + \binom{n-m + \rho_0(x)-1}{n-m}
\end{equation}
The first and last terms correspond to the number of singletons obtained by inserting either \texttt{1}'s or \texttt{0}'s, but not both. The summation over the remaining cases counts the configurations that incorporate both additional \texttt{1}'s and \texttt{0}'s.
The final result can be simplified to the identity below
\begin{equation}
   |\mathcal{S}_{n,x}| = \binom{n-m+\rho(x)_1+\rho(x)_0-1}{n-m}.
\end{equation}
\begin{theorem}
	The constant (i.e., $x=\texttt{11...1}$ or $x=\texttt{00...0}$) and the alternating $x$ strings maximize and minimize the number of singletons, respectively.
\end{theorem}
\begin{IEEEproof}
	This follows immediately from a maximization and minimization of the number of runs in $x$, i.e., $\rho_\alpha(x)$. In the case of the all \texttt{1}'s $x$ string, which comprises a single run, every index in $x$ can be used for splitting. Conversely, the alternating $x$ has the maximum number of runs $|\mathcal{R}|=m$, where $\forall. r \in \mathcal{R}_{x}: |r|=1$, thus splittings are not possible, i.e., no operations of type $(ii)$, and the insertions are confined to pre-pending and appending bits of opposite values to the first and last bit of $x$, respectively.
\end{IEEEproof}

\section{Entropy Minimization}\label{sec:entropy-minimization}
We now prove the minimal entropy conjecture for the special cases of one and two deletions. Our approach incorporates two key steps: first we work out a characterization of the number of $y$ strings that have specific weights $\omega_x(y)$. We then consider the impact of applying an entropy decreasing transformation to $x$, denoted by $g(x)$, and prove that this operation shifts the weights in the space of supersequences such that it results in a lowering of the corresponding entropy. This is achieved using clustering techniques and a run-length encoding of strings: we identify groupings of supersequences with specific weights by studying how they can be constructed from a given subsequence using different insertion operations, which are in turn based on analyzing how runs of \texttt{1}'s and \texttt{0}'s can be extended or split.

\begin{definition}\label{def:transformation-g}
We now define the transformation $g$ on strings of length $m$ as follows:
\begin{equation}
g((k_1, \dotsc, k_\ell)) =
\begin{cases}
(k_1+k_2, k_3, \dotsc, k_\ell) & \text{if $\ell > 1$} \\
g(\sigma) = \sigma
\end{cases}
\end{equation}
\end{definition}

Hence $g$ is a ``merging'' operation, that connects the two first blocks together. As we shall see, $g$ decreases the entropy. Thus, one can start from any subsequence $x$ and apply the transformation $g$ until the string becomes $\sigma$, i.e., $x=\texttt{0}^m$ or $x=\texttt{1}^m$. As a result, $\sigma$ exhibits minimal entropy and thus the highest amount of leakage in the original key exchange problem.
Note that, as indicated implicitly in the definition above, this transformation always reduces the number of runs by one by flipping the first run to its complement.

Thus we avoid cases where merging two runs would lead to connecting to a third neighboring run, thereby resulting in a reduction of runs by two. For example, $g$ transforms the string $x = \texttt{1001110} = (1;1,2,3,1)$ into $x = \texttt{0001110} = (0;3,3,1)$, as opposed to $x = \texttt{1111110} = (1;6,1)$.

\begin{figure}[tp]
	\centering
	\includegraphics[width=0.8\textwidth]{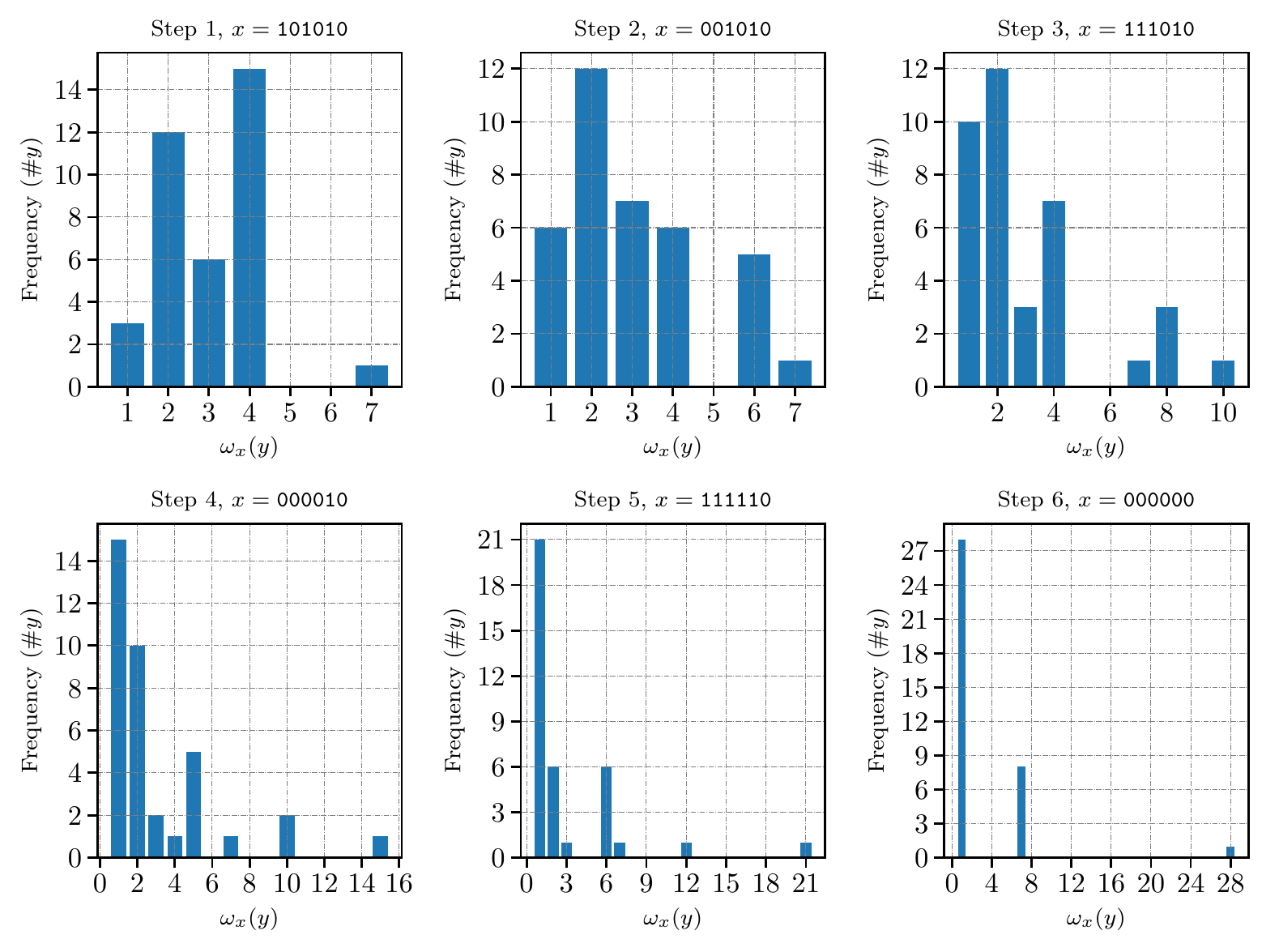}
	\caption{Impact of the transformation $g$ on the weight distribution for converting $x=\texttt{101010}$ to $x'=\texttt{000000}$, with $n=8, m=6$.}
	\label{figure:g_transform_impact}
\end{figure}

The plots shown in \Cref{figure:g_transform_impact} illustrate the impact of the transformation $g$ on the weight distribution as we move from $x=\texttt{101010}$ to $x'=\texttt{000000}$, i.e., $\texttt{101010} \rightarrow \texttt{001010} \rightarrow \texttt{111010} \rightarrow \texttt{000010} \rightarrow \texttt{111110} \rightarrow \texttt{000000}$.

\subsection{Single Deletions}\label{sec:single-deletion}

In this section we consider the case of a single deletion. Let $x$ be a fixed string of length $m$. We study the space of $y$ strings of length $n = m+1$ that can be masked to yield $x$, i.e., $Y_1 = \{y \in \{0,1\}^n \mid \exists \delta \in \mathcal{P}([n]), y_\delta = x \text{ and } |\delta| = 1\}$. Recall that we associate a weight $\omega_x(y)$ to each $y \in Y_1$, defined as the number of ways that $y$ can be masked into $x$. Finally, we define the entropy associated to $x$ as the Shannon entropy of the variable $Z \in \{0,1\}^n$ having distribution
\begin{equation*}
\Pr[Z=y] = \frac{1}{\mu_1} \omega_x(y).
\end{equation*}
where $\mu_1 = \sum_{y \in \Upsilon_{n,x}}$, which for the case $m=n-1$ gives $\binom{n}{m}2^{n-m} = \binom{n}{n-1}2^{n-(n-1)} = 2n$.

\subsubsection{Clustering Supersequences via Single Insertions}
Let $x = (k_1, \dotsc, k_\ell)$. A string $y \in Y_1$ can take only one of the following forms:
\begin{enumerate}
\item $y = (k_1, \dotsc, k_{i-1}, k_i+1, k_{i+1}, \dotsc  k_\ell) $ for some $i \in [\ell]$;
\item $y= (k_1, \dotsc, k_{i-1}, k'_i, 1, k''_i, k_{i+1}, \dotsc, k_\ell)$, for some $i \in [\ell]$ and where $k'_i + k''_i = k_i$ and $k'_i \neq 0$ and $k''_i \neq 0$;
\item $y=(1, k_1, \dotsc k_\ell)$;
\item $y= (k_1, \dotsc, k_\ell, 1)$.
\end{enumerate}
The first case will be referred to as a ``block-lengthening insertion'', denoted by $1/0$, which corresponds to extending runs/blocks. The last three cases will be referred to as ``block-splitting insertions'', and denoted by $0/1$, corresponding to splitting runs or adding a new run of length $1$. For the remainder of our discussion, $a/b$ means: ``$a$ block-lengthening insertions and $b$ block-splitting insertions''.

\begin{lemma}\label{lem:single-deletion-clustering}
$Y_1$ is composed of\footnote{A sanity check can be done to verify that we do not miss any strings, since $\binom{m+1}{m} + \binom{m+1}{m+1} = \ell + m - \ell +2$.}:
\begin{itemize}
	\item $\ell$ block-lengthening insertions, resulting in strings of respective weights $ k_1 + 1, k_2 + 1, k_3+1, \dotsc, k_\ell+1$; and
	\item $m - \ell +2$ block-splitting insertions, i.e., strings of weight 1.
\end{itemize}
\end{lemma}

\subsubsection{Minimal Entropy For Single Deletions}

\begin{lemma}\label{lemma:g-single-deletion}
The transformation $g$ decreases the entropy $H_n(x)$ for single deletions, i.e., $m=n-1$.
\end{lemma}

\begin{IEEEproof}
The proof consists of computing the difference between the entropy before and after applying $g$, i.e., $\Delta_1 = H_{n}(x) - H_n(g(x))$, and showing that this difference is positive.
From Lemma \ref{lem:single-deletion-clustering}, after applying $g$,
\begin{itemize}
\item The block-lengthening insertions give $\ell - 1$ strings of respective weights $ k_1 + 1 + k_2 + 1 -1, k_3+1, \dotsc, k_\ell+1$.
\item The block-splitting insertions give $m+2 - (\ell - 1)$ strings of weight 1.
\end{itemize}
We now compute the difference of the entropies thanks to the analyses of $(k_1, k_2, k_3, \dotsc, k_\ell)$ and $(k_1+k_2, k_3, \dotsc, k_\ell)$, which after simplification gives
\begin{equation*}
\Delta_1(k_1, \dotsc, k_\ell) = (k_1+1) \log \frac{1}{k_1+1} + (k_2+1) \log \frac{1}{k_2+1} - (k_1+k_2+1) \log \frac{1}{k_1+k_2 +1}.
\end{equation*}
This is positive, since $\log\frac{1}{k_1+1},\log\frac{1}{k_2+1} > \log\frac{1}{k_1+k_2+1}$ and $(k_1+1)+(k_2+1) > k_1+k_2+1$.
\end{IEEEproof}

\begin{corollary}\label{cor:entropy-single-deletion}
For all $n$ and any subsequence $x$ of length $m = n-1$, we have
\[H_n(x) \geq H_n\left( \sigma \right),\]
with equality only if $x\in \{\texttt{0}^m, \texttt{1}^m\}$.
\end{corollary}
\begin{IEEEproof}
Given any $x \neq \sigma$ of length $m = n-1$, it can be transformed into the string $\sigma$ by a series of consecutive $g$ operations, as defined in \Cref{def:transformation-g}. Each such operation can only decrease the entropy, as shown in Lemma \ref{lemma:g-single-deletion}, and thus we get a proof for the fact that $H_n(x) \geq H_n\left(\texttt{0}^m\right)$.
\end{IEEEproof}
\begin{remark}
It is worth pointing out that for the special case of single deletions, the minimization of entropy by the constant string, $x=[m]$, can also be proved using a simple combinatorial argument as follows. For $m=n-1$, in cluster $c=1$ we get a single $y$ string with maximum weight, $\omega_y(x)=\binom{n}{m}$, corresponding to $y=[n]$ and $x=[m]$, and the remaining strings in cluster $c=0$ are all singletons, $\omega_x(y)=1$. This is clearly the most concentrated distribution and hence the least entropic one. However our analysis shows how we will deal with the more complicated double deletions case.
\end{remark}
In the case of single deletions, we can also illustrate the utility  of our approach by deriving a stronger result for the R\'enyi entropy.

\begin{definition}[R\'enyi Entropy]
For any $\alpha>0$ and $\alpha \neq 1$, the R\'{e}nyi entropy of order $\alpha$ of a distribution $P$ is defined by
\begin{equation*}
 H_{\alpha} = \frac{1}{1-\alpha} \log_2
 \sum_{i=1}^n p_i^{\alpha}
\end{equation*}
\end{definition}

\begin{theorem}
For all $n$ and any subsequence $x$ of length $m = n-1$, and $\alpha>0$, $\alpha \neq 1$, we have that $\sigma$ exhibits the lowest R\'enyi entropy $H_\alpha$
\[H_{\alpha}(x) \geq H_{\alpha}\left( \sigma\right),\]
with equality only if $x\in \{\texttt{0}^m, \texttt{1}^m\}$.
\end{theorem}

\begin{IEEEproof}
Similar to the proof of the Shannon entropy minimization, we just have to show that $g$ decreases the R\'enyi entropy as well such that here almost all the terms also disappear and we end up with
\begin{equation*}
H_{\alpha}(x) - H_{\alpha}\left( g(x) \right) =
\frac{\alpha}{1-\alpha}\left( (k_1+1)^{\alpha} + (k_2+1)^{\alpha}  - (k_1+k_2+1)^{\alpha} \right)
\end{equation*}
This is positive since $(k_1+1)^{\alpha} + (k_2+1)^{\alpha}  - (k_1+k_2+1)^{\alpha} \geq 0 \Leftrightarrow
\alpha < 1$.
\end{IEEEproof}

\subsection{Double Deletions}

This section will follow the same structure as \Cref{sec:single-deletion}. We will enumerate the different supersequences and their corresponding weights. This is summed up in Lemma \ref{lem:double-deletion-clustering}. We then apply our analysis to a string $x$, and to its image by the function the merging operator $g$ \cref{def:transformation-g}.
\smallskip

In the case of two deletions, there are three types of insertions to consider;
using the notation introduced in the previous section, these are $2/0$, $1/1$, and $0/2$ insertions. For a fixed string $x = (k_1, \dotsc, k_\ell)$, we now analyze each case to account for the corresponding number of supersequences and their respective weights in each cluster. We will then study how this distribution changes when we go from $x$ to $g(x)$ in order to prove the following lemma:
\begin{lemma}\label{lemma:g-double-deletion}
The transformation $g$ decreases the entropy $H_n(x)$ for double deletions, i.e., $m=n-2$.

\end{lemma}

Note that while this technique could be applied to a higher number of insertions, the complexity of the analysis blows up already for two deletions, as the next section will show.

\subsubsection{Clustering Supersequences via Double Insertions}\label{sec:clustering-supersequences-double-insertion}

\paragraph{Case $2/0$}
The case $2/0$ corresponds to the situation where the insertions do not create new blocks. This happens when both bits are added to the same block, or when they are added to two different blocks, as follows.

The former corresponds to
\[
y = (k_1, \dotsc, k_{i-1},k_i+2, k_{i+1}, \dotsc, k_\ell)
\]
for some $i \in [\ell]$, which has weight $\omega_x(y) = \binom{k_i+2}{2}$. There are $\ell$ strings of this type.

The latter corresponds to
\[
y = (k_1, \dotsc, k_{i-1},k_i+1, k_{i+1}, \dotsc, k_{j-1},k_j+1, k_{j+1}, \dotsc, k_\ell)
\]
for $1 \leq i < j \leq \ell$, and has weight $\omega_x(y) = (k_i+1)(k_j+1)$. There are $\frac{\ell(\ell -1)}{2}$ strings with this weight. In total, there are $\frac{\ell(\ell+1)}{2}$ strings for the case $2/0$.

\paragraph{Case $0/2$}
In the $0/2$ case, there are only block-splitting insertions, hence all strings have weight 1. block-splitting insertions may happen in a single block, or in two separate blocks. To ease notation, we introduce
\begin{equation*}
\widetilde k_i =
\begin{cases}
k_i - 1 & \text{if $i \in [2, \ell-1]$} \\
k_i & \text{if $i = 1$ or $i = \ell$}
\end{cases}
\end{equation*}
The different treatments for ``endpoints'' $1$ and $\ell$ correspond to cases $(1, k_1, \dotsc, k_{\ell})$ and $(k_1, \dotsc, k_{\ell}, 1)$, whereas a block-splitting insertion in the $i$-th block can happen at only $k_i-1$ places.
\begin{itemize}
\item If we insert into the first or the last block, we choose respectively $k_1$ and $k_\ell$ positions, i.e., there are respectively $k_1$ and $k_\ell$ different strings.

\item If we insert into any other block $i$, we choose amongst $k_i-1$ positions, which yields $k_i-1$ different strings.
\item If we insert in different blocks, we apply the same analysis twice, independently, which gives $\widetilde{k_i}\widetilde{k_j}$ different strings.
\item If we insert twice in the same block, we get $\binom{\widetilde{k_i} + 1}{2}$ different strings.
\end{itemize}
In the end, the total number of $0/2$ insertions is
\begin{align*}
\sum_{1 \leq i < j \leq \ell}\widetilde{k_i}\widetilde{k_j} + \sum_{i=1}^\ell \binom{\widetilde{k_i} + 1}{2}
\end{align*}
\begin{example}
If $k_1 = \cdots = k_\ell = 1$, so that $\widetilde k_1 = \widetilde k_\ell = 1$ and $\widetilde k_2 = \cdots = \widetilde k_{\ell-1} = 0$, we count 3 strings.
\end{example}
\begin{example}[]
For example, for $1<i<j<l$ we get for all $a_1, a_2, b_1, b_2 > 0$ such that $a_1 + a_2 = k_i$ and $b_1 + b_2 = k_j$, the string $k_1 \dots k_{i-1} a_1 1 a_2 k_{i+1} \dots k_{j-1} b_1 1 b_2 k_{j+1} \dots k_l$. The number of such strings is $(\widetilde{k_i})(\widetilde{k_j})$.

Another example: for the particular cases $i=j=1$ we get for all $a_1,a_2,a_3 \in \mathbb{N}$ with $a_2$ strictly positive such that $a_1 + a_2 + a_3 = k_1$, the string $a_1 1 a_2 1 a_3 k_2 k_l$ or (case $a_2 = 0$) $a_1 2 a_3 k_2 k_l$. The number of such strings is $\binom{\widetilde{k_1} + 1}{2}$.
\end{example}

\paragraph{Case $1/1$}
As in the previous case, we choose a block in which we apply a block-lengthening insertion, yielding $k_i + 1$ masks; then we choose a block for a block-splitting insertion, yielding $\widetilde k_i$ strings. However, one must be careful:
to see why, consider the following string $x = \texttt{000111} = (\texttt{0};3,3)$.
\begin{itemize}
\item If we insert a block-lengthening \texttt{\good{0}} in the first block, and then a block-splitting \texttt{\bad{1}} in the last-but-one position of the first block, we get the string $y = \texttt{000\bad{1}\good{0}111} = (\texttt{0};3,1,1,3)$.
This string is of weight $(3+1) + (3+1) -1$, since we can delete the $\texttt{\good{0}}$ then one of the four $\texttt{1}$, or the $\texttt{\bad{1}}$ then one of the four $\texttt{0}$, and we remove one so that we do not double count the deletion of $\texttt{\bad{1}\good{0}}$.
\item If we insert a block-lengthening \texttt{\good{1}} in the second block, followed by a block-splitting \texttt{\bad{0}} in the second position of the second block, we obtain the same string $y = \texttt{000\good{1}\bad{0}111} = (\texttt{0}; 3,1,1,3)$.
\end{itemize}
Hence there are two ways to get each $y$. We will therefore exercise a preference toward the first situation, where we perform a block-lengthening insertion in the first block, followed by a block-splitting insertion in the first block's last-but-one position. Let $i \in [\ell]$.
\begin{itemize}
\item If $i = 1$, we get $\sum_{j=1}^{\ell}\widetilde{k_j} (=m -\ell +2)$ strings of weight $k_1 + 1$, as well as a string of weight $k_1+1+k_2$. In total, we get $m-\ell + 3$ strings.
\item If $1 < i < \ell$, we perform a block-lengthening insertion in the block $i$, the number of strings we will get is $(\sum_i \widetilde{k_i}) $. Indeed the string $(k_1, \dotsc, k_{i-1}, 1, 1, k_i, k_{i+1}, \dotsc, k_\ell)$ will be counted for the case $i-1$. Each of these strings has weight $k_i + 1 $, except one $(k_1, \dotsc, k_i, 1, 1, k_{i+1}, \dotsc, k_\ell)$ which has weight $k_i + 1 + k_{i+1}$ (the string that we will not count for $i+1$).
\item If $i=\ell$, we can keep the same formula by introducing $k_{\ell +1} = 0$ for the weight of the string $(k_1, \dotsc,  k_\ell,1,1)$.
\end{itemize}

\begin{lemma}\label{lem:double-deletion-clustering}
$Y_2$ is composed of:
\begin{itemize}
	\item case $2/0$: for all $i$ in $[\ell]$ we have one supersequence of weight $\binom{k_i+2}{2}$ and $\forall 1 \leq i < j \leq [\ell]$ we have one supersequence of weight $(k_i +1)(k_j+1)$
	\item case $0/2$: we have $\sum_{1\leq i < j \leq \ell}\widetilde{k_i}\widetilde{k_j} + \sum_{i=1}^{\ell} \binom{\widetilde{k_i}+1}{2}$ supersequences of weight $1$.
	\item case $1/1$: for all $i$ in $[\ell]$ we have $m-\ell+2$ supersequence of weight $k_i+1$ and one of weight $k_i+k_{i+1}+1$ with the convention that $k_{\ell+1}=0$.
\end{itemize}
\end{lemma}
Since the analysis is quite convoluted, we make two sanity checks on the number of supersequences and the sum of all the weights.

\begin{remark}[Sanity check for the number of supersequences]
\label{remark:numbstring:double}
We check the result of \Cref{lem:double-deletion-clustering} against \Cref{eq:upsilon-cardinality}.

We give an algebraic proof in Appendix \ref{app:2} that if $(k_i)_{i \in \{1, \dotsc, \ell \} }$ are
positive integers such that $m = \sum_{i=1}^{\ell} k_i$, then we have
\[
\frac{\ell(\ell+1)}{2} + \sum_{1 \leq i < j \leq \ell}\widetilde{k_i}\widetilde{k_j} + \sum_{i=1}^\ell \binom{\widetilde{k_i} + 1}{2} + 1 + \ell(m-\ell -2) = \binom{m+2}{m} + \binom{m+2}{m+1} + \binom{m+2}{m+2},
\]
to make sure we have not missed or double-counted any strings.
\end{remark}

\begin{remark}[Sanity check for the sum of all weights]
\label{remark:numbweight:double}
We check the result of \Cref{lem:double-deletion-clustering} against \Cref{eq:numbmask}.
Similarly, to ensure that we have not missed or double-counted any weights, we give an algebraic proof in \Cref{app:3} showing that if there exist positive integers $(k_i)_{i \in \{1, \dotsc, \ell \} }$ such that $m = \sum_{i=1}^{\ell} k_i$, then we have
\begin{align*}
 &\sum_{i=1}^{\ell}\binom{k_i+2}{2} + \sum_{1 \leq i < j \leq \ell}(k_i +1)(k_j+1) + \sum_{1\leq i < j \leq \ell}\widetilde{k_i}\widetilde{k_j} + \sum_{i=1}^{\ell} \binom{\widetilde{k_i}+1}{2} \\ &+ \sum_{i=1}^{\ell}\left[(m-\ell+2)\times (k_i+1)+k_i+k_{i+1}+1\right] = 4 \binom{m+2}{m}.
\end{align*}
\end{remark}

\subsubsection{Minimal Entropy For Double Deletions}\label{sec:double-deletion}
As in \Cref{sec:single-deletion}, we analyze the effects of the merging operation $g$ on entropy. For this, we consider the impact of $g(x) = (k_1+k_2,k_3, \dotsc, k_{\ell})$ on the clustering results developed in \Cref{sec:clustering-supersequences-double-insertion}. We will omit the analyses when no insertions are made in the first or second block, since we will get the same weight and this will disappear in the difference.

\paragraph{Case $2/0$}
For $x$, we had $\frac{\ell(\ell+1)}{2}$ strings of this type, we now have $\frac{\ell(\ell-1)}{2}$, there are $\ell$ less strings and $\ell-1$ that grow bigger. The rest remains the same.

\paragraph{Case $0/2$}
Similar to $x$, we have a certain number of strings with weight 1 counted as before \[\sum_{3 \leq i \leq j \leq \ell} \widetilde{k_i}\widetilde{k_j} + \sum_{3 \leq i \leq \ell} \binom{\widetilde{k_i} + 1}{2}\]
However, a part of the formula changes:
\begin{equation}\label{eq:1}
\binom{k_1+k_2 + 1}{2} + (k_1+k_2) \sum_{3 \leq i \leq \ell} \widetilde{k_i}
\end{equation}
Then, for the part of the analysis of $g(x)$ equivalent with that of $x$ we get
\begin{equation}\label{eq:2}
\binom{k_1 + 1}{2} + \binom{k_2}{2} + (k_1 + k_2-1) \times \sum_{3 \leq i \leq \ell} \widetilde{k_i} + k_1(k_2-1)
\end{equation}
now we take the difference between \Cref{eq:1} and \Cref{eq:2}
\begin{equation*}
\sum_{3 \leq i \leq \ell} \widetilde{k_i} + \binom{k_1 + k_2 + 1}{2} - \left(\binom{k_1 + 1}{2} + \binom{k_2}{2} + k_1(k_2-1)\right)
\end{equation*}
After simplifications, we obtain $\sum_{1 \leq i \leq \ell} \widetilde{k_i} + 1$.

\paragraph{Case $1/1$}
In the case of $x$, we had
\[(\ell-1) \sum_{1 \leq i \leq \ell}(\widetilde{k_i} - 1) + \sum_{1 \leq i \leq \ell}\widetilde{k_i}.\]
We now have
\[(\ell-2) (\sum_{1 \leq i \leq \ell}(\widetilde{k_i}-1)+1) + \sum_{1 \leq i \leq \ell}\widetilde{k_i}+1.\]
Taking the difference between now and before we get $\sum_{1 \leq i \leq \ell}{\widetilde{k_i}} + 1 - l$. We have $(\sum_{1 \leq i \leq l}(\widetilde{k_i}-1)+1)$ weights (the block-lengthening insertion in the first block) that grow bigger, the rest stays the same.

\begin{remark}[Sanity check]
We can check that the numbers of strings is constant:
\begin{itemize}
\item Case 0/2: $\sum_{1 \leq i \leq \ell} \widetilde{k_i} + 1$ more strings
\item Case 1/1: $(\sum_{1 \leq i \leq \ell}{\widetilde{k_i}} + 1 - \ell)$ less strings
\item Case 2/0: $\ell$ less strings.
\end{itemize}
and
$  \sum_{1 \leq i \leq \ell} \widetilde{k_i} + 1 -  (\sum_{1 \leq i \leq \ell}{\widetilde{k_i}} + 1 - \ell) -  \ell = 0$.
\end{remark}
We can now compute the difference of the two entropies.
Note that instead of working with the probabilities, we will multiply everything by $4\binom{m+2}{m}$ (i.e., the total number of masks). We can focus on the very few strings that show a change in weight (when an insertion is made in the first or second block).
\\
\textbf{Case $2/0$:} For $x$, we have $1$ string for each of the weights
\begin{align*}
& (k_1+1)(k_2+1), (k_1+1)(k_3+1), \dotsc, \\
& (k_1+1)(k_l+1),(k_2+1)(k_3+1), (k_2+1)(k_4+1), \dotsc, \\
& (k_2+1)(k_l+1), \binom{k_1+2}{2}\binom{k_2+2}{2}
\end{align*}
For $g(x)$, we still have $1$ string for each of the following weights:
\begin{equation*}
(k_1+k_2+1)(k_3+1), (k_1+k_2+1)(k_4+1), \dotsc, (k_1+k_2+1)(k_l+1), \binom{k_1+k_2+2}{2}
\end{equation*}
\\
\textbf{Case $0/2$:} For $g(x)$, we have $\sum_{1 \leq i \leq \ell}\widetilde{k_i}+1$.
\\
\textbf{Case $1/1$:} For $x$, the remaining strings are:
\begin{center}
\small
\begin{tabular}{cc}\toprule
Multiplicity & Weight \\\midrule
$\sum_{i=1}^{\ell}\widetilde{k_i}$ & $k_1+1$ \\
$\sum_{i=1}^{\ell}\widetilde{k_i}-1$ & $k_2+1$\\
$1$ & $k_1+k_2+1$\\
$1$ & $k_2+k_3+1$
\\\bottomrule
\end{tabular}
\end{center}
There remains, for $g(x)$, one string for each of the following weights $k_3+1,k_4+1, \dotsc, k_l+1$
and $\sum_{1 \leq i \leq \ell}\widetilde{k_i}+1$ strings of weight $k_1+k_2+1$ along with $1$ string of weight $k_1+k_2+k_3+1$. The difference of entropies is equal to the difference between $A$ and $B$ defined in the following equations:
\begin{align*}
A ={}
& \sum_{2 \leq i \leq \ell} (k_1+1)(k_i+1) \log\frac{1}{(k_1+1)(k_i+1)} +
 \binom{k_1+2}{2}\log \frac{1}{ \binom{k_1+2}{2}} + \binom{k_2+2}{2}\log \frac{1}{ \binom{k_2+2}{2}} \\
& + \sum_{1 \leq i \leq \ell}\widetilde{k_i}(k_1+1)\log\frac{1}{(k_1+1)} +(k_1+k_2+1)\log\frac{1}{(k_1 +k_2 +1)} \\
& + (\sum_{1 \leq i \leq \ell}\widetilde{k_i}-1)(k_2+1)\log\frac{1}{(k_2 +1)}\\
& + (k_2 + k_3+1)\log\frac{1}{(k_2 +k_3+1)}\\
B ={}
& \sum_{3 \leq i \leq l}(k_i+1) \log\frac{1}{(k_i+1)} + (\sum_{1 \leq i \leq l}\widetilde{k_i}+1)(k_1+k_2+1)\log\frac{1}{(k_1+k_2 +1)}\\
& + (k_1+ k_2 + k_3+1)\log\frac{1}{(k_1 + k_2 +k_3+1)}\\
&  + \sum_{3 \leq i \leq l} (k_1+k_2+1)(k_i+1) \log\frac{1}{(k_1+k_2+1)(k_i+1)} \\
& + \binom{k_1 +k_2+2}{2}\log \frac{1}{ \binom{k_1+k_2+2}{2}}
\end{align*}
where $A$ corresponds to $x$, and $B$ corresponds to $g(x)$.
We are now in a position to conclude the proof of Lemma \ref{lemma:g-double-deletion}.
\begin{lemma}\label{th:6.3}
The transformation $g$ decreases the entropy $H_n(x)$ for double deletions, i.e., $m=n-2$.
\end{lemma}
\begin{IEEEproof}
To prove this, it suffices to show that for $\ell \geq 2$, $k_i \geq 1$, $A - B > 0$. The proof mostly consists of computing partial derivatives to show that the function is increasing. We refer the reader to \Cref{app:deletions} for details.
\end{IEEEproof}

\begin{corollary}\label{thm:entropy-two-deletions}
For all $n$ and any subsequence $x$ of length $m = n-2$, we have
\[H_n(x) \geq H_n\left( \sigma \right),\]
with equality only if $x\in \{\texttt{0}^m, \texttt{1}^m\}$.
\end{corollary}
\begin{IEEEproof}
Given any $x \neq \sigma$ of length $m = n - 2$, it can be transformed into the string $\sigma$ by a series of consecutive $g$ operations (cf. \Cref{def:transformation-g}). Each such operation can only decrease the entropy, as proved in Lemma \ref{lemma:g-double-deletion}, and thus we get a proof for the fact that $H_n(x) \geq H_n\left(\texttt{0}^m\right)$.
\end{IEEEproof}

\section{Concluding Remarks}\label{sec:conclusion}

From the original cryptographic motivation of the problem, the minimal entropy case corresponding to maximal information leakage is arguably the case that interests us the most. While our results shed more light on various properties of the space of supersequences and the combinatorial problem of counting the number of embeddings of a given subsequence in the set of its compatible supersequences, the original entropy maximization conjecture remains an open problem. Finally, proving the entropy minimization conjecture for an arbitrary number of deletions as well as a more general characterization of the distribution of the number of subsequence embeddings in supersequences of finite-length present some further open problems.

\appendices
\section{Proof of Lemma \ref{th:6.3}}\label{app:deletions}
\begin{IEEEproof}
The proof consists of two steps: first we show that $A-B > 0$ for all $k_1 \geq 1$ when $k_2 = \cdots = k_\ell = 1$; then we show that $\nabla(A-B)$ is positive along all directions others than the first one, so that an increase in any of the $k_i$ with $i\geq 2$, results in an increase of $A-B$.
We start by simplifying the expression.
To do so, we introduce the function $e(x) = -x \log_2 x$. We also use the fact that $e(xy) = xe(y)+ye(x)$, and develop the binomial coefficients: $e\left(\binom{a+b}{2} \right) = \binom{a+b}{2} + e((a+b)(a+b-1))$. Then we match the sum indexes. We also introduce the notation $e_i = e(k_i + 1)$.

Thus we can write:
\begin{align*}
A ={}
& \sum_{2 \leq i \leq l} e((k_1+1)(k_i+1)) +
 e\left(\binom{k_1+2}{2}\right) + e\left(\binom{k_2+2}{2}\right)\\
& +e(k_1+1) \sum_{1 \leq i \leq l}\widetilde{k_i}+e(k_1+k_2+1) \\
& + e(k_2+1) \sum_{1 \leq i \leq l}(\widetilde{k_i}-1)\\
& + e(k_2 + k_3+1) \\
={}
& \sum_{3 \leq i \leq \ell-1} e((k_1+1)(k_i+1)) + e((k_1+1)(k_2+1)) + e((k_1+1)(k_\ell+1)) \\
& + \binom{k_1+1}{2} + e((k_1+1)(k_1+2))\\
& + \binom{k_2+1}{2} + e((k_2+1)(k_2+2))\\
& +e(k_1+1) \sum_{3 \leq i \leq \ell-1}\widetilde{k_i}+e(k_1+k_2+1) + e(k_1+1)\widetilde{k_1} + e(k_1+1)\widetilde{k_2} + e(k_1+1)\widetilde{k_\ell}\\
& + e(k_2+1) \sum_{3 \leq i \leq \ell- 1}\widetilde{k_i}- \ell e(k_2+1) + e(k_2+1)\widetilde{k_1} + e(k_2+1)\widetilde{k_2} + e(k_2+1)\widetilde{k_\ell}\\
& + e(k_2 + k_3+1)\\
={}
& (k_1+1)\sum_{3 \leq i \leq \ell-1}e_i + e_1\sum_{3 \leq i \leq \ell-1}(k_i+1)  \\
& + (k_2+1)e_1 + (k_1+1)e_2 + (k_1+1)e_\ell + (k_\ell+1)e_1\\
& + \binom{k_1+1}{2} + (k_1+2)e_1 + (k_1+1)e(k_1+2)\\
& + \binom{k_2+1}{2} + (k_2+2)e_2 + (k_2+1)e(k_2+2) \\
& +e_1 \sum_{3 \leq i \leq \ell-1} k_i  - (\ell-3)e_1 +e(k_1+k_2+1) + e_1 k_1 + e_1 k_2 - e_1 + e_1 k_\ell\\
& + e_2 \sum_{3 \leq i \leq \ell- 1} k_i - (\ell-3)e_2 - \ell e_2 + e_2 k_1 + e_2 k_2 - e_2 + e_2 k_\ell\\
& + e(k_2 + k_3+1)
\end{align*}
At this point we regroup all terms in $e_i$ together:
\begin{align*}
A ={}
& \left( 2k_1 + 2k_2 + 2k_\ell - 3 + 2\sum_{3 \leq i \leq \ell-1} k_i \right)e_1 \\
& + \left( 2k_1 + 2k_2 + k_\ell - 2\ell - 1 + \sum_{3 \leq i \leq \ell- 1} k_i \right)e_2\\
& + (k_1+1)\sum_{3 \leq i \leq \ell-1}e_i \\
& + (k_1+1)e_\ell \\
& + e(k_1+k_2+1) + (k_1+1)e(k_1+2) + (k_2+1)e(k_2+2) + e(k_2 + k_3+1) \\
& + \binom{k_1+1}{2} + \binom{k_2+1}{2}
\end{align*}
We simplify the expression for $B$ in the same fashion:
\begin{align*}
B ={}
& 2e(k_1+k_2+1)\sum_{3 \leq i \leq \ell-1}k_i + k_1 e(k_1+k_2+1) + k_2 e(k_1+k_2+1) + k_\ell e(k_1+k_2+1) \\
& + e(k_1+ k_2 + k_3+1)\\
& +  (\ell - 3)e(k_1+k_2+1) + (k_1+k_2+2) \sum_{3 \leq i \leq \ell} e_i\\
& + \binom{k_1 +k_2+2}{2} + e((k_1+k_2+1)(k_1+k_2+2)) \\
={}
& (k_1+k_2+2) \sum_{3 \leq i \leq \ell-1} e_i  \\
& +\left(2k_1 + 2k_2 + k_\ell + \ell - 1 + 2\sum_{3 \leq i \leq \ell-1}k_i \right)e(k_1+k_2+1)\\
& + e(k_1+ k_2 + k_3+1) +  (k_1+k_2+1)e(k_1+k_2+2) + (k_1+k_2+2)e_\ell\\
& + \binom{k_1 +k_2+2}{2} \\
\end{align*}
so that we can now compute the difference:
\begin{align*}
A - B
={}
& \left( - 3 + 2\sum_{1 \leq i \leq \ell} k_i \right)e_1 \\
& + \left( k_1 + k_2 - 2\ell - 1 + \sum_{1 \leq i \leq \ell} k_i \right)e_2\\
& - (k_2+1)\sum_{3 \leq i \leq \ell}e_i  \\
& + (k_1+1)e(k_1+2) + (k_2+1)e(k_2+2) \\
& + 1 - k_1 k_2 \\
& -\left(-k_\ell + \ell + 2\sum_{1 \leq i \leq \ell}k_i \right)e(k_1+k_2+1)\\
& - (k_1+k_2-1)e(k_1+k_2+2) - e(k_1+ k_2 + k_3+1)  + e(k_2 + k_3+1) \\
={}
& P(\vec k)e_1 + Q(\vec k)e_2 - (k_2+1)\sum_{i=3}^{\ell} e_i + (k_1+1)e(k_1+2) + (k_2+1)e(k_2+2) \\
& + 1 - k_1k_2 - R(\vec k)e(k_1+k_2+1) - (k_1+k_2-1)e(k_1+k_2+2) - e(k_1+ k_2 + k_3+1) \\
& + e(k_2 + k_3+1).
\end{align*}
Where
\begin{align*}
P(\vec k) & = - 3 + 2\sum_{1 \leq i \leq \ell} k_i,\\
Q(\vec k) & =  k_1 + k_2 - 2\ell - 1 + \sum_{1 \leq i \leq \ell}k_i \\
R(\vec k) & = -k_\ell + \ell + 2\sum_{1 \leq i \leq \ell}k_i .
\end{align*}
We now compute $A-B$ where $k_i=1$ for $i \geq 2$ and show that it is positive. We get:
\begin{align*}
& (3 \ell - 3 + 2k_1)(k_1+2) \log_2 (k_1+2) +
(k_1+1)(k_1+3)\log_2 (k_1+3) + 4(\ell - 2)(\log_2 2) +1 \\
& - [ (2k_1+2\ell -5)(k_1+1)\log_2 (k_1+1) +
(k_1+1)(k_1+2)\log_2 (k_1+2) + 2(2k_1 - \ell -1)\log_2 2 + 9 \log_2 3 + k_1] \\
& = (2k_1 + 2 \ell -5)(k_1 + 1) \log (k_1+1) [\log_2 (k_1+2) -\log_2 (k_1+2) ] +
(3 \ell - 3 + 2 k_1 + (\ell + 2)(k_1+2)) \log_2 (k_1+2) \\
& + (k_1+1)(k_1+2)[ \log_2 (k_1+2) - \log_2 (k_1+1) ] + (k_1+1) \log_2 (k_1+3) \\
& + 4(\ell - 2)(\log_2 2) +1 - [ 2(2k_1 - \ell -1)\log_2 2 + 9 \log_2 3 + k_1]
\end{align*}
Since $k_1 \geq 1$ and $\ell \geq 2$ we have $2k_1 \log_2 (k_1+2) + (k_1+1) \log_2 (k_1+3) \geq 2(2k_1 - \ell -1)\log_2 2$, $3 (\ell-1)\log_2 (k_1+2) + 2(k_1+2) \log_2 (k_1+2) \geq 9 \log_2 3$ and $\ell (k_1+2) \log_2 (k_1+2) \geq k_1$. This suffices to conclude that $A-B$ is positive when $k_i=1$ for $i\geq2$.\\
We now compute the partial derivatives for $i\geq2$ and show that they are positive.
The gradient can be computed term by term thanks to linearity, observing that for any polynomial $S(\vec k)$,
\begin{align*}
\partial_i e_i & = -\log_2(k_i+1) - \frac{1}{\ln(2)} \\
\partial_i e_j & = 0 \qquad (i \neq j)\\
\nabla S(\vec k)e_j & = \left(e_j\partial_i S(\vec k)+ S(\vec k)\partial_i e_j \right)_{i=1}^{\ell}
\end{align*}
Hence, by denoting $\vec u_1, \dotsc, \vec u_\ell$ the canonical basis, we have:
\begin{align*}
\nabla P(\vec k)e_1 & =  \left(e_1\partial_i P(\vec k)+ P(\vec k)\partial_i e_1 \right)_{i=1}^{\ell} = \partial_1 e_1 P(\vec k) \vec u_1 + e_1(\partial_i P(\vec k))_{i=1}^{\ell} \\
& = \partial_1 e_1 P(\vec k) \vec u_1 + 2(\vec u_1 + \cdots + \vec u_\ell) \\
& = (2 + \partial_1 e_1 P(\vec k))\vec u_1 + 2\vec u_2 + \cdots + 2\vec u_\ell \\
\nabla Q(\vec k)e_2 & = \left(e_2\partial_i Q(\vec k)+ Q(\vec k)\partial_i e_2 \right)_{i=1}^{\ell} = \partial_2 e_2 Q(\vec k)\vec u_2 + (\partial_i S(\vec k))_{i=1}^{\ell} \\
& = \partial_2 e_2 Q(\vec k)\vec u_2 + \vec u_1 + \vec u_2 + \vec u_1 + \cdots + \vec u_\ell \\
& = 2\vec u_1 + (2 + \partial_2 e_2 Q(\vec k))\vec u_2 + \vec u_3 + \cdots + \vec u_\ell \\
-\nabla \left((k_2+1)\sum_{i=3}^\ell e_i \right) & =
-(k_2+1)\nabla\sum_{i=3}^\ell e_i -(\nabla(k_2+1))\sum_{i=3}^\ell e_i \\
& = -((k_2+1)\partial_i e_i \vec u_i)_{i=3}^{\ell}-\left(\sum_{i=3}^{\ell}e_i\right)\vec u_2 \\
\nabla \left( (k_j+1)e(k_j+2)\right) & = -\left( \log_2(k_j+2) + \frac{1}{\ln(2)}\frac{k_j+1}{k_j+2}\right)\vec u_j \\
\nabla(1 - k_1k_2) & = -k_2\vec u_1 -k_1 \vec u_2 \\
- \nabla \left( R(\vec k)e(k_1+k_2+1) \right)
& = -R(\vec k) \nabla e(k_1+k_2+1) - e(k_1+k_2+1)\nabla R(\vec k) \\
& = R(\vec k) \left(\log_2(k_1+k_2+1) + \frac{1}{\ln(2)} \right)(\vec u_1 + \vec u_2) \\
& \qquad - e(k_1+k_2+1)(\partial_i R(\vec k))_{i=1}^{\ell} \\
& = R(\vec k) \left(\log_2(k_1+k_2+1) + \frac{1}{\ln(2)} \right)(\vec u_1 + \vec u_2) \\
& \qquad - e(k_1+k_2+1)(2\vec u_1 + \cdots + 2\vec u_{\ell-1} + \vec u_\ell) \\
- \nabla (k_1+k_2-1)e(k_1+k_2+2) & = -(k_1+k_2-1)\nabla e(k_1+k_2+2) - e(k_1+k_2+2) \nabla (k_1+k_2-1) \\
& = (k_1+k_2-1)\left( \log_2(k_1+k_2+2) + \frac{1}{\ln(2)} \right)(\vec u_1 + \vec u_2)  \\
& \qquad - e(k_1+k_2+2)(\vec u_1 + \vec u_2) \\
& = \left((k_1+k_2-1)\left( \log_2(k_1+k_2+2) + \frac{1}{\ln(2)} \right) - e(k_1+k_2+2)\right)(\vec u_1 + \vec u_2)\\
- \nabla e(k_1+ k_2 + k_3+1) & = \left(\log_2(k_1 + k_2 + k_3 + 1) + \frac{1}{\ln(2)}\right)(\vec u_1 + \vec u_2 + \vec u_3) \\
\nabla e(k_2 + k_3+1) & = -\left( \log_2(k_2+k_3+1) + \frac{1}{\ln(2)}\right)(\vec u_2 + \vec u_3)
\end{align*}
As is clearly visible from the above equations, we only need to consider the components along $\vec u_2$, $\vec u_3$, $\vec u_\ell$, and along $\vec u_i$ for any $3 < i < \ell$. For the latter, we have
\begin{align*}
\left( \nabla (A-B)\right)_i
& = 2 + 1 -(k_2+1)\partial_i e_i - 2e(k_1 + k_2 + 1) \\
& = 3 + 2(k_1 + k_2 + 1)\log_2(k_1 + k_2 + 1) +(k_2+1) \left(\log_2(k_i + 1) + \frac{1}{\ln(2)} \right) \\
& > 0.
\end{align*}
Now, along the very similar $\vec u_\ell$ axis,
\begin{align*}
\left( \nabla (A-B)\right)_\ell
& = 2 + 1 - (k_2+1)\partial_\ell e_\ell -e(k_1+k_2+1) \\
& = 3 + (k_1 + k_2 + 1)\log_2(k_1 + k_2 + 1) + (k_2+1) \left(\log_2(k_\ell + 1) + \frac{1}{\ln(2)} \right) \\
& > 0.
\end{align*}
Along $\vec u_3$,
\begin{align*}
\left( \nabla (A-B)\right)_3
& = 2 + 1 - (k_2+1)\partial_3 e_3 - 2e(k_1+k_2+1) + \log_2(k_1+k_2+k_3+1) + \frac{1}{\ln(2)} \\
& \qquad - \log_2(k_2+k_3+1) + \frac{1}{\ln(2)} \\
& = 3 + 2(k_1 + k_2 + 1)\log_2(k_1 + k_2 + 1) + (k_2+1) \left(\log_2(k_3 + 1) + \frac{1}{\ln(2)} \right) \\
& \qquad + \log_2(k_1+k_2+k_3+1) - \log_2(k_2+k_3+1) \\
& > 0.
\end{align*}
Along $\vec u_2$,
\begin{align*}
\left( \nabla (A-B)\right)_2
={} & 2 + 2 + Q(\vec k) \partial_2 e_2 - (k_2+1)\partial_2 e_2 \\
& - \sum_{i=3}^\ell e_i -\left(\log_2(k_2+2) + \frac{1}{\ln(2)}\frac{k_2+1}{k_2+2} \right) - k_1 \\
&  + R(\vec k)\left( \log_2(k_1+k_2+1) + \frac{1}{\ln(2)} \right) - 2e(k_1+k_2+1) \\
&  + \left((k_1+ k_2-1)\left( \log_2(k_1+k_2+2) + \frac{1}{\ln(2)}\right) -e(k_1+k_2+2) \right) \\
&  + \log_2(k_1+k_2+k_3+1) + \frac{1}{\ln(2)} - \log_2(k_2+k_3+1) - \frac{1}{\ln(2)}  \\
={} & 4 - (Q(\vec k) - k_2-1) \left( \log_2(k_2+1) + \frac{1}{\ln(2)}\right)   - \frac{1}{\ln(2)}\frac{k_2+1}{k_2+2} - k_1 \\
&  + \sum_{i=3}^\ell (k_i+1)\log_2(k_i + 1) \\
&  + R(\vec k)\left( \log_2(k_1+k_2+1) + \frac{1}{\ln(2)} \right) - 2e(k_1+k_2+1) \\
&  + (k_1+ k_2)\left( \log_2(k_1+k_2+2) + \frac{1}{\ln(2)}\right) -e(k_1+k_2+2)  \\
&  + \log_2(k_1+k_2+k_3+2) - \log_2(k_2+k_3+2) -\log_2(k_2+2) \\
\end{align*}
\begin{lemma}\label{lem:Q1}
$\left( \nabla (A-B)\right)_2 > 0$.
\end{lemma}
\begin{IEEEproof}[Proof of \Cref{lem:Q1}]
Letting $\lambda = \frac{1}{\ln(2)}$
We first show that the following line is positive
\begin{align*}
& -(Q (\vec k) - k_2 -1)(\log_2(k_2+1) + \lambda) - \lambda \frac{k_2+1}{k_2+2} - k_1 +  \\
& + R(\vec k)(\lambda + \log_2(k_1+k_2+1)) \\
={} & \lambda \left(R(\vec k) - Q(\vec k) + k_2 + 1 - \frac{k_2+1}{k_2+2}\right) + R(\vec k)\log_2(k_1+k_2+1) \\
&  - Q(\vec k)\log_2(k_2+1) \\
={} & \lambda \left( \sum_{i=2}^{\ell-1} k_i + 3\ell + 1 - \frac{k_2+1}{k_2+2}\right) \\
& + R(\vec k)\log_2(k_1+k_2+1) - Q(\vec k)\log_2(k_2+1).
\end{align*}
The last line is positive since in particular $ R(\vec k)\log_2(k_1+k_2+1) - Q(\vec k)\log_2(k_2+1) > (R(\vec k) - Q(\vec(k))\log_2(k_2+1)>0$.
Note that $-e(k_1+k_2+2)-\log_2(k_2+2) > 0$, $\log_2(k_1+k_2+k_3+1)-\log_2(k_2+k_3+1)>0$ and the remaining quantities are positive.
\end{IEEEproof}
As a result, we have that $A - B > 0$ for all $\vec k$ such that $k_i \geq 1$, which establishes the theorem.
\end{IEEEproof}

\section{Proof of Remark \ref{remark:numbstring:double}}\label{app:2}
We prove that for all positive integer sequences $(k_i)_{i \in \{1, \dotsc , \ell\}}$ such that
$\sum_{i=1}^{\ell}k_i = m$ we have :
\begin{align*}
  \frac{\ell(\ell+1)}{2} + \sum_{1 \leq i < j \leq \ell}\widetilde{k_i}\widetilde{k_j} + \sum_{i=1}^\ell \binom{\widetilde{k_i} + 1}{2} + 1 + \ell(m-\ell -2)
  = \binom{m+2}{m} + \binom{m+2}{m+1} + \binom{m+2}{m+2}
\end{align*}
We fix $\ell$ and $m$, then proceed by induction on the sequences of $(k_i)_{i \in \{1, \dotsc , \ell\}}$. We first show the equality for $k_1 = m- \ell +1$, and $k_i = 1$ for all $i> 1$.
\begin{IEEEproof}
We have on the left hand side:
\begin{align*}
&\frac{\ell(\ell+1)}{2} + \sum_{1 \leq i < j \leq \ell}\widetilde{k_i}\widetilde{k_j} + \sum_{i=1}^\ell \binom{\widetilde{k_i} + 1}{2} + 1 + \ell(m-\ell -2) \\
& = \frac{\ell(\ell+1)}{2} + 1 + \ell(m-\ell -2) + (m- \ell + 1) \sum_{j=2}^{\ell} \widetilde{k_j} + \binom{2}{2} + \binom{m-\ell+2}{2} \\
& = \frac{1}{2}\left(\ell (\ell +1) + (m-\ell+2)(m-\ell+1)\right) + (\ell +1)(m-\ell+2) + 1 \\
& = \frac{1}{2}(m^2 + 3m + 2) + m + 3 \\
& = \binom{m+2}{m} + \binom{m+2}{m+1} + \binom{m+2}{m+2}
\end{align*}
which concludes the initialization.
\end{IEEEproof}
We now fix a sequence $(k_i)_{i \in \{1,\dotsc , \ell\}}$, and $i_0 \in \{1, \dotsc , \ell\}$. We assume that the equality holds for this sequence and show that it is true for the sequence $(k'_i)_{i \in \{1, \dotsc , \ell\}}$ defined as $k'_i = k_i$ if $i \neq i_0$ and $i \neq i_0 + 1$, $k'_{i_0} = k_{i_0} - 1$ and $k'_{i_0+1} = k_{i_0+1} + 1$.
\begin{IEEEproof}
We first note that only a part of the formula on the left hand side depends on $(k_i)_{i \in \{1, \dotsc , \ell\}}$. Letting
\[
F\left((k_i)_{i \in \{1, \dotsc , \ell \}}\right) =  \sum_{1 \leq i < j \leq \ell}\widetilde{k_i}\widetilde{k_j} + \sum_{i=1}^\ell \binom{\widetilde{k_i} + 1}{2},
\]
we just have to prove that
\begin{equation*}
F\left((k_i)_{i \in \{1, \dotsc , \ell\}}\right) - F\left((k'_i)_{i \in \{1, \dotsc , \ell\}}\right) = 0.
\end{equation*}
Expanding the above difference, we have:
\begin{align*}
&(\widetilde{k_{i_0}} - \widetilde{k'_{i_0}})(\sum_{i=1}^{i_0-1}\widetilde{k_i})
+  (\widetilde{k_{i_0 + 1}} - \widetilde{k'_{i_0 + 1}})(\sum_{i=1}^{i_0-1}\widetilde{k_i}) + \widetilde{k_{i_0}} \widetilde{k_{i_0+1}} - \widetilde{k'_{i_0}}\widetilde{k'_{i_0+1}}
\\
& \qquad +
(\widetilde{k_{i_0}} - \widetilde{k'_{i_0}})(\sum_{i=i_0+1}^{\ell}\widetilde{k_i})
+  (\widetilde{k_{i_0 + 1}} - \widetilde{k'_{i_0 + 1}})(\sum_{i=i_0+2}^{\ell}\widetilde{k_i}) \\
& \qquad  + \binom{\widetilde{k_{i_0}}+1}{2} - \binom{\widetilde{k'_{i_0}}+1}{2} + \binom{\widetilde{k_{i_0+1}}+1}{2} - \binom{\widetilde{k_{i_0+1}}+1}{2}
\end{align*}
This is equal to
\begin{align*}
& \widetilde{k_{i_0}} \widetilde{k_{i_0+1}} - (\widetilde{k_{i_0}}-1)(\widetilde{k_{i_0+1}}+1) + \widetilde{k_{i_0+1}} + \frac{1}{2}\Big(\widetilde{k_{i_0}} (\widetilde{k_{i_0}}+1) - (\widetilde{k_{i_0}} +1)\widetilde{k_{i_0}}\Big) \\
& \qquad - \frac{1}{2}\Big(\widetilde{k_{i_0+1}} (\widetilde{k_{i_0+1}}+1) - (\widetilde{k_{i_0+1}} + 2)\widetilde{k_{i_0+1}}+1\Big) \\
& = - \widetilde{k_{i_0}} + \widetilde{k_{i_0+1}} + 1 + \frac{1}{2}(2\widetilde{k_{i_0}} - 2 \widetilde{k_{i_0+1}} - 2) \\
&= 0.
\end{align*}
This concludes the proof.
\end{IEEEproof}

\section{Proof of Remark \ref{remark:numbweight:double}}\label{app:3}
As in \Cref{app:2} we proceed by induction to show that
if there exist positive integers $(k_i)_{i \in \{1, \dotsc, \ell \} }$ such that $m = \sum_{i=1}^{\ell} k_i$, then we have
\begin{align*}
\sum_{i=1}^{\ell}&\binom{k_i+2}{2} + \sum_{1 \leq i < j \leq \ell}(k_i +1)(k_j+1) + \sum_{1\leq i < j \leq \ell}\widetilde{k_i}\widetilde{k_j} + \sum_{i=1}^{\ell} \binom{\widetilde{k_i}+1}{2} \\+ &\sum_{i=1}^{\ell}\left[(m-\ell+2)\times (k_i+1)+k_i+k_{i+1}+1\right] = 4 \binom{m+2}{m}
\end{align*}

We fix $\ell$ and $m$, then proceed by induction on the sequences of $(k_i)_{i \in \{1, \dotsc , \ell\}}$. We first show the equality for $k_1 = m- \ell +1$, and $k_i = 1$ for all $i> 1$.
\begin{IEEEproof}
We have on the left hand side of the equation:
\begin{align*}
& (m-\ell+1)+\binom{m-\ell+2}{2}+1+\binom{m-\ell+3}{2}+(\ell-1)*3+(\ell-1)(m-\ell+2)*2+2(\ell-2)(\ell-1)+(m-\ell+2)^2 \\
& +m-\ell+3+(\ell-1)(2(m-\ell+1)+3)-1 \\
& = 2m^2 + 6 m + 4 \\
& = 4 \binom{m+2}{m}
\end{align*}
which concludes the initialization.
\end{IEEEproof}
We now fix a sequence $(k_i)_{i \in \{1,\dotsc , \ell\}}$, and $i_0 \in \{1, \dotsc , \ell\}$. We assume that the equality holds for this sequence and show that it is true for the sequence $(k'_i)_{i \in \{1, \dotsc , \ell\}}$ defined as $k'_i = k_i$ if $i \neq i_0$ and $i \neq i_0 + 1$, $k'_{i_0} = k_{i_0} - 1$ and $k'_{i_0+1} = k_{i_0+1} + 1$.
\begin{IEEEproof}
First notice that as in \Cref{app:2} we can ignore all the terms that do not depends on the $k_i$. Furthermore we can reuse the result of \Cref{app:2} to remove
$\sum_{1 \leq i < j \leq \ell}\widetilde{k_i}\widetilde{k_j} + \sum_{i=1}^\ell \binom{\widetilde{k_i} + 1}{2}$.
We also note that
\[
\sum_{i=1}^{\ell}\left[(m-\ell+2)\times (k_i+1)+k_i+k_{i+1}+1\right] =
\sum_{i=1}^{\ell}\left[(m-\ell+2)\times (k'_i+1)+k'_i+k'_{i+1}+1\right]
\]
since $\sum_{i=1}^{\ell} k_i = \sum_{i=1}^{\ell}k'_i$. We therefore define
\[
F\left((k_i)_{i \in \{1, \dotsc , \ell \}}\right) = \sum_{i=1}^{\ell}\binom{k_i+2}{2} + \sum_{1 \leq i < j \leq \ell}(k_i +1)(k_j+1)
\]
and show that
\begin{equation*}
F\left((k_i)_{i \in \{1, \dotsc , \ell\}}\right) - F\left((k'_i)_{i \in \{1, \dotsc , \ell\}}\right) = 0.
\end{equation*}
Expanding the difference we get
\begin{align*}
 &\binom{k_{i_0}+2}{2} - \binom{k'_{i_0}+2}{2}  + \binom{k_{i_0+1}+2}{2} - \binom{k'_{i_0+1}+2}{2} \\
&  + k_{i_0}\sum_{j>i_0}^{\ell}(k_j+1) - k'_{i_0}\sum_{j>i_0}^{\ell}(k'_j+1)
+ k_{i_0+1}\sum_{j>i_0+1}^{\ell}(k_j+1) - k'_{i_0+1}\sum_{j>i_0+1}^{\ell}(k'_j+1) \\
& = k_{i_0}+1 - (k_{i_0+1} + 2) + k_{i_0}S - (k_{i_0}-1) \left(S+1 \right) +
k_{i_0+1} S' - (k_{i_0+1} +1) S' \\
& = 1 - k_{i_0} + S - 1 - S' \\
& = 0
\end{align*}
where $S = \sum_{j>i_0}^{\ell}(k_j+1)$ and $S'=\sum_{j>i_0+1}^{\ell}(k_j+1)$. This concludes the proof.
\end{IEEEproof}

\newpage

\bibliographystyle{IEEEtran}
\bibliography{IEEEabrv,main}

\end{document}